\documentclass[a4paper]{article}

\usepackage{epsfig}
\usepackage{amssymb,amsmath,amsfonts}
\usepackage{multirow}
\usepackage{rotating}
\usepackage{graphicx}
\usepackage{lscape}
\usepackage{hyperref}
\usepackage{ulem}

\usepackage{cite} 

\usepackage{geometry}
\newcommand{\GRACE}{\textsc{G\hspace{-.1em}r\hspace{-.1em}a\hspace{-.1em}ce}}

\newcommand{\SARAH}{\textsc{S\hspace{-.1em}a\hspace{-.1em}rah}}
\def\g2{{g\hspace{-.2em}-\hspace{-.2em}2}}

\begin{document}
\title{
Two Higgs doublet model fitting and $t\bar{t}b\bar{b}$ signal at the ILC
}
\author{Hieu Minh Tran$^{a)}$, Huong Thu Nguyen$^{b)}$, and Yoshimasa Kurihara$^{c)}$\\ \\
a) \textit{Hanoi University of Science and Technology,}\\
\textit{1 Dai Co Viet Road, Hanoi, Vietnam}\\
b) \textit{Univ of Science, Vietnam National University, Hanoi,}\\
\textit{334 Nguyen Trai Road, Hanoi, Vietnam}\\
c) \textit{The High Energy Accelerator Organisation (KEK),}\\
\textit{Tsukuba, Ibaraki 305-0801, Japan}
}
\maketitle
\begin{abstract} 
One of the most straightforward extensions of the standard model (SM) is having an additional Higgs doublet to the SM, namely the two Higgs doublet models(THDM).
In the type-I model, an additional Higgs doublet is introduced that
does not couple to any fermion via the Yukawa interaction
in the original Lagrangian.
Considering various theoretical and phenomenological constraints, we have found the best-fitted parameter set in the type-I model using the minimum $\chi^2$ method by scanning the model's parameter space.
We show that this optimal parameter set can be tested by precisely analyzing the decay processes 
$D^+_s \rightarrow \tau^+ \nu_\tau$,
$D^+ \rightarrow \mu^+ \nu_\mu$, and
$B^0 \rightarrow K^* \mu^+ \mu^-$.
Moreover, the decay channels 
$h \rightarrow \gamma\gamma$ and $Z\gamma$ can be used to distinguish the model from the SM.
For a direct search of the model at future colliders, we have proposed an investigation of the $e^+e^- \rightarrow t\bar{t} b\bar{b}$ process at the ILC to detect the new physics of the model.
Considering the initial state radiation correction and applying appropriate background cuts, the calculation result of the scattering cross section shows that it is feasible to observe a clear and unique signal in the $t\bar{b}$ invariant mass distribution corresponding to the charged Higgs pair creations.

\end{abstract}
\maketitle
\newpage
%
%
%
\section{Introduction}
In the standard model (SM), the minimal one that can generate the fermion masses after the spontaneous symmetry breaking, the scalar sector consists of only one Higgs doublet.
Since scalars are so flexible for model building, they have been used in various new physics models to address theoretical challenges and experimental anomalies.
Although the SM-like Higgs boson was discovered at the LHC, there may be still more Higgs bosons in nature that come from an extended scalar sector.
Among various extensions of the SM, the two Higgs doublet model (THDM)\cite{Haber:1978jt} is an important one since its structure can be found in many new physics models such as the minimal supersymmetric extension of the SM \cite{Martin:1997ns, Okada:2010xe, Tran:2010ea}, grand unified theories\cite{Croon:2019kpe, Fukuyama:2019zun}, composite Higgs models, axion models\cite{Peccei:1977hh, Peccei:1977ur}.
Moreover, the additional CP violation from the scalar potential of the THDM helps realize the successful electroweak baryogenesis\cite{Turok:1990in, Nelson:1991ab, Funakubo:1993jg, Cline:1995dg,Cline:1996mga, Laine:2000rm, Fromme:2006cm, Davies:1994id}.

The THDM generally suffers from the flavour-changing neutral current (FCNC) problem.
The two Yukawa coupling matrices relevant to the Higgs doublets in the model can not be diagonalized simultaneously by the rotational matrices relating the initial fermion states and their mass eigenstates, unlike the situation in the SM.
This leads to the dangerous tree-level FCNCs that are strictly constrained by experiments.
To ameliorate this problem, a $Z_2$ symmetry was introduced in the theory\cite{Branco:2011iw} so that a given fermion couples to only one Higgs doublet, as dictated by the theorem of Glashow and Weinberg \cite{Glashow:1976nt}.
Another way to avoid the FCNCs is to assume the flavour alignment of the Yukawa couplings.
According to that, the Yukawa matrices of the two scalar doublets are proportional\cite{Manohar:2006ga, Pich:2009sp, Penuelas:2017ikk}.
The free parameter space of the THDM has been investigated, taking into account various constraints, including the LEP and LHC data, flavour observables and theoretical requirements\cite{Jung:2010ik, Chen:2013kt, Celis:2013rcs, Chiang:2013ixa, Grinstein:2013npa, Eberhardt:2013uba, Celis:2013ixa, Chang:2013ona, Wang:2013sha, Baglio:2014nea, Ilisie:2014hea, Kanemura:2014bqa, Bernon:2014nxa, Craig:2015jba, Abbas:2015cua, Botella:2015hoa, Ilnicka:2015jba, Bernon:2015qea, Bernon:2015wef, Cacchio:2016qyh, Belusca-Maito:2016dqe, Ilnicka:2018def, Dercks:2018wch, Botella:2018gzy, Sanyal:2019xcp, Herrero-Garcia:2019mcy, Karmakar:2019vnq, Chen:2019pkq, Arco:2020ucn, Aiko:2020ksl, Botella:2020xzf, Athron:2021auq, Botella:2022rte, Botella:2023tiw}
The global fit to the THDM with the soft $Z_2$-symmetry breaking was performed in Refs. \cite{Chowdhury:2015yja, Chowdhury:2017aav, Haller:2018nnx}.
For the case of the aligned THDM, the global fit was carried out in Refs. \cite{Eberhardt:2020dat, Karan:2023kyj}.
It has been shown that searches for the THDM signatures in future colliders are promising
\cite{Kling:2018xud, Adhikary:2018ise,Kon:2018vmv, Arco:2021bvf, Han:2021udl,Li:2020hao, Liu:2020kxt, Chung:2022kjp,Bahl:2020kwe}.

Depending on the specific form of the Lagrangian, the THDM is usually classified into four separated types (I, II, X, and Y) \cite{Aoki:2009ha}.
In each of these models, the scalar sector interacts differently with the fermion sector, resulting in different interesting phenomenologies.
In this paper, we focus our analysis on the THDM of type I, where the fermions only couple to one scalar doublet, while the other does not couple to any fermion in the original Lagrangian.
Therefore, all of the fermion masses are generated by the vacuum expectation value of that scalar doublet.
We perform the data fitting of the model to various theoretical and phenomenological constraints, taking into account the results of the LEP, the Tevatron, and the LHC experiments, the measurements of flavour observables, and the current world-average value of the $W$-boson mass.
For the best-fit point, the rare decays of $B$ and $D$ mesons, as well as the properties of the SM-like Higgs boson are analyzed in detailed.
The result indicates that the decay processes 
$D^+ \rightarrow \mu^+ \nu_\mu$, 
$B^0 \rightarrow K^* \mu^+ \mu^-$,
$h \rightarrow s\bar{s}$, $W^+W^-$ and $gg$ will be sensitive to the new physics in the model.
These are the clues to probe the model with more precise data in the future.
To resemble the experimental results consistent with the SM predictions, couplings of the heavy Higgs particles to fermions are very small. 
Therefore, it is challenging to distinguish between the SM and the THDM by comparing the total cross sections.
Nevertheless, looking at the invariant mass distributions after appropriate cuts, in principal we can see typical peaks to identify the corresponding new particles.
Since the Yukawa couplings of Higgs bosons to heavier fermions are larger, processes with top and bottom quarks in the final state are expected to be more sensitive to the heavy Higgs bosons than other channels.
Using the best-fit point, we examine the four-fermion-production channel
$e^+ e^- \rightarrow t\, \bar{t} \, b \, \bar{b}$ and investigate the possibility of detecting the new physics signals at the International Linear Collider (ILC).

The structure of the paper is as follows.
In Section 2, we briefly review the type-I THDM.
In Section 3, the data fitting is performed to find the best-fit point of the model.
Here, we also examine the rare decays of $B$ and $D$ mesons, and the branching ratios of the SM-like Higgs boson to find the suggestion of new physics implication in current data.
The $e^+e^-$ scattering analysis is performed in Section 4 to look for possible signals of the model at the ILC.
Finally, Section 5 is devoted to summary and discussion.

\section{Two Higgs doublet model}
In the THDM, two scalar doublets are introduced to generate masses for fermions and gauge bosons after acquiring their vacuum expectation values. Its Lagrangian is given by
\begin{align}
\mathcal{L}_\text{THDM}	&=
	\mathcal{L}_\text{SM}^\text{no scalar} 
	+ \mathcal{L}_\text{scalar}
	+ \mathcal{L}_\text{Yukawa},
\end{align}
where $\mathcal{L}_\text{SM}^\text{no scalar}$ involves the SM terms without any scalar particles,
$\mathcal{L}_\text{scalar}$ is the Lagrangian for the two scalar doublets including their kinetic terms with covariant derivatives and the scalar potential $\mathcal{V}_\text{scalar}$,
and 
$\mathcal{L}_\text{Yukawa}$ is the one describing the Yukawa interactions between fermions and scalars.
To avoid the dangerous FCNCs emerging from the Yukawa Lagrangian, a $Z_2$ symmetry is usually imposed in the theory. 
Under this discrete symmetry, the two scalars transform differently, i.e. 
one is even
($\Phi_1 \rightarrow \Phi_1$)
while the other is odd
($\Phi_2 \rightarrow -\Phi_2$) 
by convention.
Therefore, for any choice of 
$Z_2$-transformation rules for the fermions that keeps $\mathcal{L}_\text{Yukawa}$ invariant 
\cite{Aoki:2009ha}, a given fermion can couple to only one scalar field:
\begin{align}
\mathcal{L}_\text{Yukawa}	&=
	- \bar{Q}_L Y_u \tilde{\Phi}_u u_R
	- \bar{Q}_L Y_d \Phi_d d_R
	- \bar{L}_L Y_\ell \Phi_\ell \ell_R
	+ \text{h.c.} ,
\label{Yukawa}
\end{align}
where $\Phi_{u,d,\ell}$ is either $\Phi_1$ or $\Phi_2$.
Depending on the  $Z_2$ charges of the fermions, the THDM are categorized into four distinct types, so-called I, II, X, and Y.
For example, in the type-I THDM, the left(right)-handed fermions are $Z_2$ even (odd).
Hence, the scalar doublet $\Phi_2$ couples to all of the fermions and generates their mases after the spontaneous breaking of the electroweak symmetry, while $\Phi_1$ does not
(namely, $\Phi_u = \Phi_d = \Phi_\ell = \Phi_2$ in Eq. (\ref{Yukawa})).

For the scalar potential of 
$\Phi_1$ and $\Phi_2$, we allow the soft $Z_2$ breaking term so that the discrete symmetry can be restored in the UV limit
\cite{Ginzburg:2004vp}.
The scalar potential is given by
\begin{align}
V(\Phi_1, \Phi_2) &=
	m_{11}^2 \Phi_1^\dagger \Phi_1
	+ m_{22}^2 \Phi_2^\dagger \Phi_2
	- \left( 
		m_{12}^2 \Phi_1^\dagger \Phi_2 + \text{h.c.}
	  \right)\notag \\
&	+ \frac{\lambda_1}{2} (\Phi_1^\dagger \Phi_1)^2
	+ \frac{\lambda_2}{2} (\Phi_2^\dagger \Phi_2)^2
	+ \lambda_3 (\Phi_1^\dagger \Phi_1) (\Phi_2^\dagger \Phi_2) 
	+ \lambda_4 |\Phi_1^\dagger \Phi_2|^2
	+ \left(
		\frac{\lambda_5}{2} (\Phi_1^\dagger \Phi_2)^2 + \text{h.c.}
	  \right).
\end{align}
%
%
%
%
%
Denoting the vacuum expectation values (VEVs) of the neutral components of the two scalar doublets as 
$\langle \Phi_{1,2}^0 \rangle = { v_{1,2}}/{\sqrt{2}}$, 
the minimization equations,
${\partial V}/{\partial \Phi^0_i} |_{\Phi^0_i = \langle \Phi_i^0 \rangle} = 0$, 
of the scalar potential allow us to represent $m_{11}^2$ and $m_{22}^2$ in terms of other parameters
\cite{Barroso:2013awa, Aoki:2021oez}:
\begin{align}
m_{11}^2		&=
	m_{12}^2 \frac{v_2}{v_1} 
	- \frac{\lambda_1}{2} v_1^2
	- \frac{1}{2} (\lambda_3 + \lambda_4 + \lambda_5) v_2^2	 ,	\\
m_{22}^2		&=
	m_{12}^2 \frac{v_1}{v_2} 
	- \frac{\lambda_1}{2} v_1^2
	- \frac{1}{2} (\lambda_3 + \lambda_4 + \lambda_5) v_2^2	 .
\end{align}

It is convenient to work in the Higgs basis $(\phi, \phi')$ where only one of them ($\phi$) acquires a nonzero VEV,
$\langle \phi^0 \rangle = \frac{v}{\sqrt{2}}$,
where
$v = \sqrt{v_1^2 + v_2^2}$.
The transformation between the two bases reads
\begin{align}
\left(
	\begin{matrix}
	\Phi_1	\\ \Phi_2
	\end{matrix}
\right) 
&=
\left(
	\begin{matrix}
	\cos \beta	&	-\sin\beta	\\ 
	\sin \beta	&	\cos\beta
	\end{matrix}
\right)
\left(
	\begin{matrix}
	\phi		\\ \phi'
	\end{matrix}
\right) ,
\end{align}
where $\tan\beta={v_2}/{v_1}$.
The components of the scalar doublets, 
$\phi$ and $\phi'$, are
\begin{equation}
\phi	 = 
\left(
	\begin{matrix}
	G^+	\\
	\frac{h_1' + v + iG^0}{\sqrt{2}}
	\end{matrix}
\right) , \quad
\phi' = 
\left(
	\begin{matrix}
	H^+	\\
	\frac{h_2' + iA}{\sqrt{2}}
	\end{matrix}
\right) ,
\end{equation}
where $(H^\pm, G^\pm)$ and $(A, G^0)$
are the mass eigenstates.
In the Higgs basis,
the massive components ($H^\pm$ and $A$) are identified as the charged Higgs boson and the CP-odd  Higgs,
while the massless components ($G^\pm$ and $G^0$) are the are the Nambu-Goldstone bosons to be absorbed by the $W^\pm$ and $Z$ gauge bosons.
The mass eigenstates of the CP-even Higgs are obtained by another rotation:
\begin{align}
\left(
	\begin{matrix}
	h'_1	\\
	h'_2
	\end{matrix}
\right)	&=
	\left(
		\begin{matrix}
		\cos(\beta-\alpha)	& \sin(\beta-\alpha)	\\
		-\sin(\beta-\alpha)	&	\cos(\beta-\alpha)
		\end{matrix}
	\right)
	\left(
		\begin{matrix}
		H	\\
		h
	\end{matrix}
	\right) .
\end{align}
Hereafter, we utilize abbreviations $s_\bullet:=\sin\bullet$ and $c_\bullet:=\cos\bullet$.
The masses of the physical states 
($m_{h,H,A,H^\pm}$) 
and the mixing angles ($\alpha$, $\beta$) are related to the couplings $\lambda_{1,...,5}$ in the scalar potential as
\cite{Barroso:2013awa, Chakraborty:2015raa}
\begin{align}
\lambda_1	&=
	\frac{1}{v^2 c^2_\beta}
	\left(
		c^2_\alpha m_H^2 
		+ s^2_\alpha m_h^2
		- m_{12}^2 \frac{s_\beta}{c_\beta}
	\right) ,	\\
\lambda_2	&=
	\frac{1}{v^2 s^2_\beta}
	\left(
		s^2_\alpha m_H^2 
		+ c^2_\alpha m_h^2
		- m_{12}^2 \frac{c_\beta}{s_\beta}
	\right) ,	\\
\lambda_3	&=
	\frac{2 m_{H^\pm}^2}{v^2} 
	+ \frac{s_{2\alpha}}{v^2 s_{2\beta}} (m_H^2 - m_h^2) 
	-\frac{m_{12}^2}{v^2 s_\beta c_\beta} 		,	\\
\lambda_4	&=
	\frac{m_A^2 - 2m_{H^\pm}^2}{v^2}
	+ \frac{m_{12}^2}{v^2 s_\beta c_\beta}
	, \\
\lambda_5	&=
	\frac{m_{12}^2}{v^2 s_\beta c_\beta}
	- \frac{m_A^2}{v^2} .
\end{align}
Identifying $h$ as the SM-like Higgs boson with $m_h = 125.20$ GeV
\cite{ParticleDataGroup:2024},
and the SM VEV 
$v = (\sqrt{2} G_F)^{1/2}  \approx 246$ GeV,
where $G_F$ is the Fermi constant,
the scalar potential is then determined by six independent real parameters:
\begin{align}
m_H , \,  m_A, \, m_{H^\pm}, \, m_{12}^2, \, 
\tan\beta , \,
\sin(\beta-\alpha).
\end{align}

\section{Parameter optimization}
We consider the theoretical constraints, including the scalar potential's stability, the S-matrix's tree-level unitarity\cite{Ginzburg:2005dt}, and the perturbativity.
The phenomenological constraints are those on the oblique parameters ($S$, $T$, $U$) and the various flavour observables related to the decay of $B$ and $D$ mesons.
Furthermore, we consider the constraint given by the world-averaging value of the $W$ boson mass the Particle Data Group\cite{ParticleDataGroup:2024}.
For this purpose, the theoretically predicted value for $m_W$ in the THDM with the loop contributions is given as \cite{Peskin:1991sw}
\begin{align}
m_W	&=	
	\sqrt{
		\left(m_W^\text{SM}\right)^2 +
		\frac{\alpha \cos^2{\hspace{-.2em}\theta_W}}{\cos^2{\hspace{-.2em}\theta_W} - \sin^2{\hspace{-.2em}\theta_W}}
		m_Z^2
		\left(
			- \frac{1}{2} S
			+ \cos^2{\hspace{-.2em}\theta_W} T
			+ \frac{\cos^2{\hspace{-.2em}\theta_W} - \sin^2{\hspace{-.2em}\theta_W}}{4 \sin^2{\hspace{-.2em}\theta_W}} U
		\right)
	} \, ,
\end{align}
where $m_W^\text{SM}$, $\theta_W$and $\alpha$ are the SM predicted value for the $W$ boson mass, the Weinberg mixing angle   and the fine structure constant, respectively.
The experimental data for these constraints are given in Table \ref{Constraints}.

For the numerical analysis, we use the 2HDMC package version 1.8.0 \cite{Eriksson:2009ws}
to examine the theoretical constraints, 
and to calculate the oblique parameters.
The flavor observables are calculated by the SuperIso package version 4.1 \cite{Mahmoudi:2007vz, Mahmoudi:2008tp, Mahmoudi:2009zz}.
The HiggsBounds function \cite{Bechtle:2008jh, Bechtle:2011sb, Bechtle:2012lvg, Bechtle:2013wla, Bechtle:2020pkv, Bahl:2021yhk}
of the HiggsTools package \cite{Bahl:2022igd}
is used to check if a parameter point of the model is consistent with the Higgs searches at the LEP, the Tevatron and the LHC.
\begin{table}[h!]
\begin{center}
\begin{tabular}{c||c|c}
Observables	&	Exp. data	& [Refs.]	\\
\hline
S & $ -0.04 \pm 0.10$	& \cite{ParticleDataGroup:2024}	\\
T & $ 0.01 \pm 0.12$	&	\cite{ParticleDataGroup:2024}	\\
U & $ -0.01 \pm 0.09 $	&	\cite{ParticleDataGroup:2024}	\\
\hline
BR($b\rightarrow s\gamma$) & $(3.49 \pm 0.19)\times 10^{-4}$	&	\cite{HeavyFlavorAveragingGroup:2022wzx}	\\
BR($B_s^0 \rightarrow \mu^+ \mu^-$) & $(3.34 \pm 0.27)\times 10^{-9}$	&	\cite{ParticleDataGroup:2024}		\\
BR($B_d^0 \rightarrow \mu^+ \mu^-$) & $(0.12 \pm 0.08)\times 10^{-9}$	&	\cite{LHCb:2021vsc}	\\
BR($B_u^+ \rightarrow \tau^+ \nu_\tau$)	&	$(1.09 \pm 0.24)\times10^{-4}$	& \cite{ParticleDataGroup:2024}	\\
BR($D_s^+ \rightarrow \tau^+ \nu_\tau$)	&	
$(5.36 \pm 0.10)\times 10^{-2}$	& \cite{ParticleDataGroup:2024}	\\
BR($D_s^+ \rightarrow \mu^+ \nu_\mu$)	&	
$(5.35 \pm 0.12)\times 10^{-3}$	& \cite{ParticleDataGroup:2024}	\\
BR($D^+ \rightarrow \mu^+ \nu_\mu$)	&	$(3.74 \pm 0.17)\times 10^{-4}$	& \cite{ParticleDataGroup:2024}		\\
BR($B^+ \rightarrow \bar{D}^0 \tau^+ \nu_\tau$)	&	$(7.7 \pm 2.5)\times 10^{-3}$		& \cite{ParticleDataGroup:2024}		\\
$\dfrac{\text{BR}(B \rightarrow D \tau \nu_\tau)}{\text{BR}(B \rightarrow D \ell \nu_\ell)}$	&	$0.339 \pm 0.030$	&	\cite{HeavyFlavorAveragingGroup:2022wzx}	\\
BR$(\bar{B} \rightarrow X_s \ell^+ \ell^-)_\text{low}$	&	$(1.58 \pm 0.37)\times 10^{-6}$	&	\cite{Huber:2015sra, Belle-II:2018jsg, Huber:2024rbw}	\\
BR$(\bar{B} \rightarrow X_s \ell^+ \ell^-)_\text{high}$	&	$ (3.23 \pm 0.47) \times 10^{-7}$	&	\cite{Huber:2015sra, Belle-II:2018jsg, Huber:2024rbw}	\\
BR$(B^0 \rightarrow K^* \mu^+ \mu^-)_\text{low}$	&	$(1.68 \pm 0.15)\times 10^{-7}$	&	\cite{LHCb:2016ykl}	\\
BR$(B^0 \rightarrow K^* \mu^+ \mu^-)_\text{high}$	&	$(1.74 \pm 0.14)\times 10^{-7}$	&	\cite{LHCb:2016ykl}	\\
\hline
$m_W$ (GeV)	&	$ 80.369 \pm 0.013 $	& \cite{ParticleDataGroup:2024}	\\
\end{tabular}
\end{center}
\caption{List of observables and experimental data used in the fitting.}
\label{Constraints}
\end{table}

Focusing on the case that the lightest CP-even Higgs is the SM-like Higgs boson,
the ranges of the independent input parameters are chosen as follows
\begin{align}
0.01 &< \tan \beta < 100,	\\
125 \text{ GeV} &< m_H, m_A, m_{H^\pm} < 2000 \text{ GeV},	\\
0 &< \sin (\beta - \alpha) < 1,	\\
-10^8 \text{ GeV}^2 &< m_{12}^2 < 10^8 \text{ GeV}^2 .
\end{align}
For each parameter point, we determine the 
value of the $\chi^2$ function defined as follows
\begin{align}
\chi^2_\text{Total}	&=
	\chi^2_\text{HS} + \sum_n \chi^2_n  ,
\label{chi2}
\end{align}
where $\chi^2_\text{HS}$ is calculated by the HiggsSignals function 
\cite{Bechtle:2013xfa, Stal:2013hwa, Bechtle:2014ewa, Bechtle:2020uwn} 
of the HiggsTools package, 
and the sum in the second term is taken over all the considered observables in Table \ref{Constraints}.
The best fit point is the one minimizing the above $\chi^2$ function.

\begin{figure}[h!]
	\begin{center}
	\includegraphics[scale=0.8]{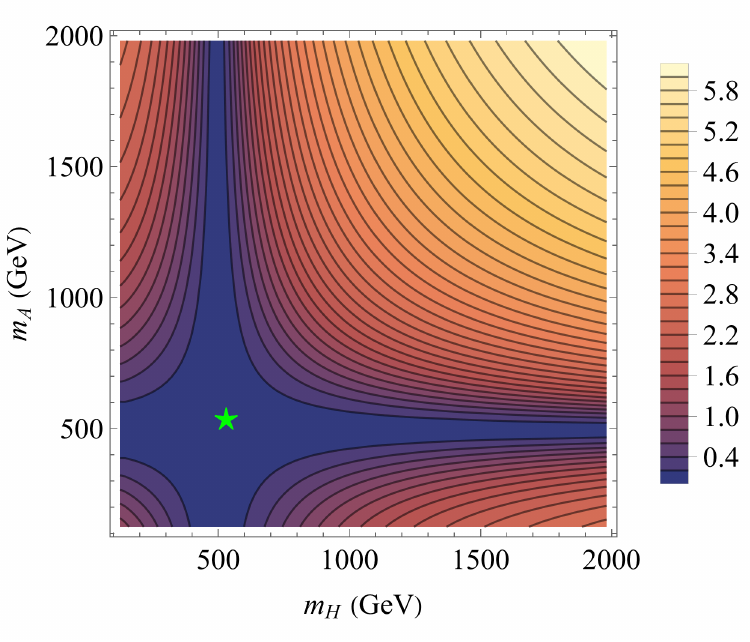}
	\caption{
	Contour plot of $\text{ln}\left(\frac{\chi^2}{N}\right)$ in the parameter space of $(m_H, m_A)$. Here, $N=176$ is the number of observables in the fitting. The green star represents the best-fit point.}	
	\label{mHmA}
	\end{center}
\end{figure}

To achieve this goal, we randomly scan over the parameter space with the MultiNest package version 3.11\cite{Feroz:2007kg, Feroz:2008xx, Feroz:2013hea}, where the nested sampling algorithm is used.
An optimum parameter point minimizing the $\chi^2$ function in Eq. (\ref{chi2}) is obtained after the convergence criterion is satisfied.
This scan is carried out many times to avoid falling into some local minimum.
Subsequently, the smallest minimum found in the above steps is refined further using the Nelder-Mead simplex algorithm \cite{Nelder:1965zz, 10.2307/2346772} with a high precision.
As a result, we obtained the best-fit parameter set shown in Table 
\ref{Bestfitpoint}.
In the above fitting procedure, the $\chi^2$ value is demanded to be at the precision of $10^{-8}$.
Therefore, the best-fit point appears to be relatively fine-tuned.
It is worth mentioning that this tuning due to the required precision is significantly different from the fine-tuning regarding to the gauge hierarchy (or the naturalness) problem.
\begin{table}[h!]
\begin{center}
\begin{tabular}{c|c}
Parameter & Best-fit value \\
\hline
$\tan\beta$	&	66.86936665	\\
$m_H$		&	530.4203175 GeV	\\
$m_A$		&	536.4124109 GeV	\\
$m_{H^\pm}$	&	493.9057661 GeV	\\
$\sin (\beta -\alpha)$	& 0.99999999975	\\
$m_{12}^2$	&	4.205745992 $\times 10^3$ GeV$^2$	\\
\end{tabular}
\end{center}
\caption{The best-fit parameter set.}
\label{Bestfitpoint}
\end{table}
%
In Figure \ref{mHmA}, this point is showed in the parameter space of $(m_H, m_A)$ as the green star. 
We see that the best-fit point stays in the middle of the intersection area between two flat directions.

\begin{figure}[h!]
	\begin{center}
		\includegraphics[scale=1]{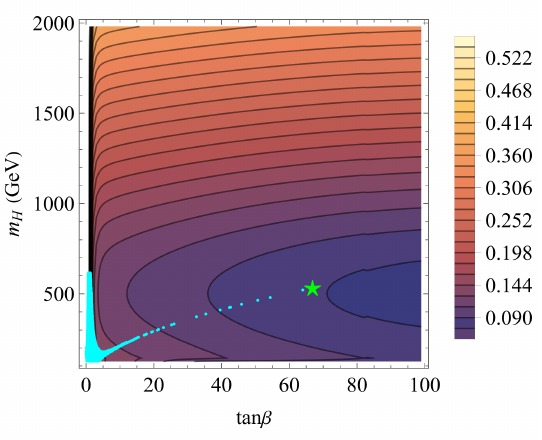}
		\caption{
			Contour plot of $\text{ln}\left(\frac{\chi^2}{N}\right)$ in the parameter space of $(\tan\beta,m_{H})$, where $N=176$ is the number of observables in the fitting. The green star represents the best-fit point.
			The thin cyan strip is allowed by the perturbativity and the stability constraints.
			}
		\label{tanbmH}
	\end{center}
\end{figure}

\begin{figure}[h!]
	\begin{center}
		\includegraphics[scale=1]{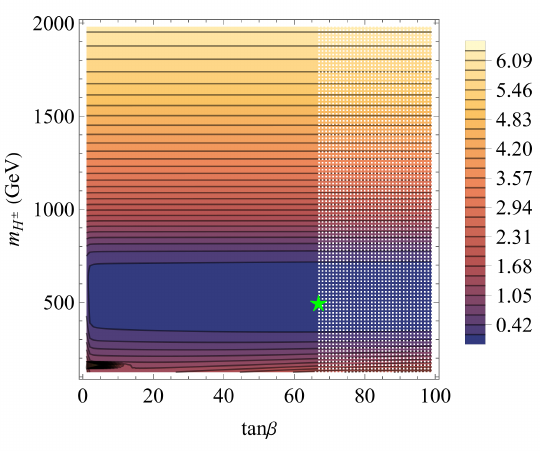}
		\caption{
		Contour plot of $\text{ln}\left(\frac{\chi^2}{N}\right)$ in the parameter space of $(\tan\beta,m_{H^\pm})$, where $N=176$ is the number of observables in the fitting. The green star represents the best-fit point.
		The dotted region is excluded by the stability constraint.
		}
		\label{tanbmHp}
	\end{center}
\end{figure}

In Figure \ref{tanbmH}, the contour plot of the 
$\ln \left(\frac{\chi^2}{N}\right)$ function is shown in the $(\tan\beta,m_H)$ plane.
Fixing $\tan\beta$, the best-fit point locates at the minimum $\chi^2$ value along the direction of $m_H$. 
When fixing $m_H$ and increasing $\tan\beta$, the $\chi^2$ function becomes smaller due to the suppression of the relevant new physics contributions. 
However, once $\tan\beta$ is larger than the best-fit value, the combined constraint imposed by perturbativity and stability is violated. As we see in Figure \ref{tanbmH}, the best-fit point stays at the tip of the thin cyan strip allowed by these two constraints. 
The similar contour plot on the $(\tan\beta,m_{H^\pm})$ plane is depicted in Figure \ref{tanbmHp}. The dotted region is excluded by the stability constraint.
According to Ref. \cite{Enomoto:2015wbn}, the branching ratios in Table \ref{Constraints} tend to push up $m_{H^\pm}$ and $\tan\beta$.
In Figure \ref{tanbmHp}, we see that $\tan\beta$ cannot exceed the best-fit value (the green star) in order to preserve the stability condition.
Moreover, we find that too large $m_{H^\pm}$ would result in the inconsistency between the theoretical prediction and the measurement of the oblique parameter $T$ \cite{Grimus:2008nb}.
Therefore, taking into account all the considered constraints, the preferred range for $m_{H^\pm}$ is around 500 GeV.
When fixing the value of $\tan\beta$, the best-fit value of $m_{H^\pm}$ corresponds to the minimum of the $\chi^2$ function.

\begin{table}[h!]
\begin{center}
\begin{tabular}{c||c|c||c}
Observables	&	THDM prediction	& Pulls ($\sigma$)	&	SM prediction	\\
\hline
S & 3.84$\times 10^{-3}$	& $0.44$	&	0 \\
T & 2.76$\times 10^{-2}$	& $0.15$	&	0 \\
U & 9.69$\times 10^{-5}$	& $0.11$	&	0 \\
\hline
BR($b\rightarrow s\gamma$) & 3.18	$\times 10^{-4}$& $-1.64$	&	3.40$\times 10^{-4}$	\\
BR($B_s^0 \rightarrow \mu^+ \mu^-$) & 2.96$\times 10^{-9}$	& $-1.41$	&	3.25$\times 10^{-9}$	\\
BR($B_d^0 \rightarrow \mu^+ \mu^-$) & 9.45$\times 10^{-11}$	& $-0.32$	&	9.63$\times 10^{-11}$	\\
BR($B_u^+ \rightarrow \tau^+ \nu_\tau$)	&	8.37$\times 10^{-5}$	& $-1.05$ 	&	8.69$\times 10^{-5}$	\\
BR($D_s^+ \rightarrow \tau^+ \nu_\tau$)	&	
5.12$\times 10^{-2}$	& $-2.37$	&	5.22$\times 10^{-2}$	\\
BR($D_s^+ \rightarrow \mu^+ \nu_\mu$)	&	
5.26$\times 10^{-3}$	& $-0.77$	&	5.36$\times 10^{-3}$	\\
BR($D^+ \rightarrow \mu^+ \nu_\mu$)	&	4.14$\times 10^{-4}$	& $2.37$	&	4.01$\times 10^{-4}$	\\
BR($B^+ \rightarrow \bar{D}^0 \tau^+ \nu_\tau$)	&	6.77$\times 10^{-3}$	& $-0.37$	&	6.90$\times 10^{-3}$	\\
$\dfrac{\text{BR}(B \rightarrow D \tau \nu_\tau)}{\text{BR}(B \rightarrow D \ell \nu_\ell)}$	&	0.315
	& $-0.81$	&0.301	\\
BR$(\bar{B} \rightarrow X_s \ell^+ \ell^-)_\text{low}$	&	1.41$\times 10^{-6}$		& $-0.45$	&	1.73$\times 10^{-6}$	 \\
BR$(\bar{B} \rightarrow X_s \ell^+ \ell^-)_\text{high}$	&	4.12$\times 10^{-7}$		& $1.89$	&	3.21$\times 10^{-7}$	\\
BR$(B^0 \rightarrow K^* \mu^+ \mu^-)_\text{low}$	&	2.09$\times 10^{-7}$		& $2.70$	&	2.48$\times 10^{-7}$	\\
BR$(B^0 \rightarrow K^* \mu^+ \mu^-)_\text{high}$	&	1.98$\times 10^{-7}$		& $1.73$	&	2.21$\times 10^{-7}$	\\
\hline
$m_W$ (GeV)	&	80.353 & $-1.23$	&	80.353	\\
\end{tabular}
\end{center}
\caption{The theoretical predictions in the THDM and pulls of physical observables for the best-fit point with $m_h = 125.20$ GeV \cite{ParticleDataGroup:2024}.
The corresponding SM values are shown in the last column
\cite{ParticleDataGroup:2024,
	Misiak:2020vlo,
	Charles:2004jd,
	CKMFitterLink,
	Chen:2006nua,
	Huber:2024rbw, 
	Descotes-Genon:2015uva}
.}
\label{Theoretical_values}
\end{table}

In Table \ref{Theoretical_values}, the theoretical values of the observables listed in Table \ref{Constraints} are presented for the case of the best-fit point.
The pulls in terms of the standard deviation for each observable are indicated in the last column of this table.
We observe that the pulls for other observables are all smaller than $2\sigma$, except for the pulls corresponding to the branching ratios of the decay processes 
$D_s^+ \rightarrow \tau^+ \nu_\tau$, $D^+ \rightarrow \mu^+ \nu_\mu$ and $B^0 \rightarrow K^* \mu^+ \mu^-$ at low momentum transfer, that are 
-2.37,
2.37 and  2.70 respectively.
Therefore, these three decay processes will be the sensitive probes to test the best-fit point in the future when the experimental precision is increased.
In the last column of Table \ref{Theoretical_values}, the SM predictions of the corresponding observables are shown.
By comparing the THDM and the SM values of the favor observables, we observe that their differences are of $\mathcal{O}(1)\%$ in most cases except for the processes
$b\rightarrow s \gamma$,
$B_s^0 \rightarrow \mu^+ \mu^-$,
$\bar{B} \rightarrow X_s \ell^+ \ell^-$, and
$B^0 \rightarrow K^* \mu^+\mu^-$ whose deviations are of $\mathcal{O}(10)\%$.
In general, the processes involving heavier quarks are more sensitive to new physics.
In the type-I THDM, there are additional contributions to the $b\rightarrow s$ transitions from the loop diagram involving the charged Higgs boson.
It is worth noticing that the theoretical uncertainties in the calculations of flavor observables significantly contribute to the deviations between the predictions of the two models. Therefore, we assume that the predicted values in our analysis still hold in more precise evaluations.

\begin{table}[h!]
\begin{center}
\begin{tabular}{c||c|c|c||c|c}
Observables	&	THDM 	& SM		& 
$\delta(\%)$		&	Exp. data	&	Dev. $(\sigma)$\\
\hline \hline
BR($h\rightarrow s \bar{s}$)	& 2.161$\times10^{-4}$	&	2.097$\times 10^{-4}$	&	3.0	&		&		\\
BR($h\rightarrow c \bar{c}$)	& 3.171$\times10^{-2}$	&	3.133$\times10^{-2}$	&	1.2
	&		&		\\
BR($h\rightarrow b \bar{b}$)	& 5.793$\times10^{-1}$	&	5.751$\times10^{-1}$	&	0.7
		& 	$(5.3 \pm 0.8)\times 10^{-1}$	&	0.62		\\
BR($h\rightarrow \mu \bar{\mu}$)	& 2.157$\times 10^{-4}$	&	2.163$\times 10^{-4}$	&	-0.3
	& 	$(2.6 \pm 1.3)\times 10^{-4}$	& 	$-$0.34	\\
BR($h\rightarrow \tau \bar{\tau}$)	& 6.228$\times 10^{-2}$	&	6.236$\times 10^{-2}$	&	-0.1
	& 	$(6.0^{+0.8}_{-0.7})\times 10^{-2}$	&	0.33		\\
BR($h\rightarrow \gamma \gamma$)	&	2.409$\times 10^{-3}$	&	2.269$\times 10^{-3}$	&	6.2
	& 	$(2.50 \pm 0.20)\times 10^{-3}$	& 		$-$0.45	\\
BR($h\rightarrow Z Z$)	& 2.698$\times 10^{-2}$	&		2.674$\times 10^{-2}$	&	0.9	& 	$(2.80 \pm 0.30) \times 10^{-2}$	& 	$-$0.34	\\
BR($h\rightarrow W^+ W^-$)	&	2.160$\times 10^{-1}$	&	2.185$\times 10^{-1}$	&	-1.1
	&	$(2.57 \pm 0.25) \times 10^{-1}$	&	$-$1.64	\\
BR($h\rightarrow Z \gamma$)	&	1.641$\times 10^{-3}$	&	1.544$\times 10^{-3}$	&	6.3	& 	$(3.4 \pm 1.1)\times 10^{-3}$	& 	$-$1.60	\\
BR($h\rightarrow g g$)	&	7.921$\times 10^{-2}$	&	8.170$\times 10^{-2}$	&	-3.1	&		&	\\
\hline
$\Gamma^h_\text{total}$ (GeV)	&	4.085$\times 10^{-3}$	&	4.117$\times 10^{-3}$	&	-0.8
	& 	$(3.7^{+1.9}_{-1.4})\times 10^{-3}$	& 	0.20		\\
\end{tabular}
\end{center}
\caption{
The branching ratios and the total decay width of the light Higgs boson ($m_h = 125.20$ GeV  \cite{ParticleDataGroup:2024}) for the best-fit point (2nd column)
in the THDM in comparison with the SM predictions\cite{LHCHiggsCrossSectionWorkingGroup:2016ypw, LHCHiggsCrossSectionWorkingGroup:2013rie
	} (3rd column), and the experimental data \cite{ParticleDataGroup:2024} (5th column).
The 4th column ($\delta$) shows the relative differences between the THDM values and the SM predictions.
The 6th column indicates the deviations between the THDM values and the experimental data.
}
\label{h-BR}
\end{table}

The properties of the SM-like Higgs boson including the branching ratios and the total decay width have been calculated for the best-fit point of the THDM 
using the package 2HDECAY \cite{Krause:2018wmo}.
The results are shown in the second column of Table \ref{h-BR}.
For comparison, the corresponding SM predicted values \cite{LHCHiggsCrossSectionWorkingGroup:2016ypw, LHCHiggsCrossSectionWorkingGroup:2013rie} shown in  the third column have been computed using the package HDECAY \cite{Djouadi:1997yw, Djouadi:2018xqq}.
	In these calculations, both the QCD and the electroweak corrections
	\cite{Krause:2016oke, Krause:2016xku, Denner:2018opp, Hahn:1998yk}
	to the decay widths have been considered. The necessary inputs for the calculations have been updated according to Ref. \cite{ParticleDataGroup:2024}.
	The relative difference between the branching ratios of each decay channel ``$i$'' predicted by two models, shown in the fourth column, can be decomposed as
\begin{align}
		\delta_i \equiv 
		\frac{\text{BR}_i^\text{THDM} - \text{BR}_i^\text{SM}}{\text{BR}_i^\text{SM}} 
		=
		\frac{\Gamma_i^\text{THDM}}{\Gamma_\text{total}^\text{THDM}} \times
		\frac{\Gamma_\text{total}^\text{SM}}{\Gamma_i^\text{SM}} - 1
		\simeq
		\frac{\Delta\Gamma_i^\text{NP}}{\Gamma_i^\text{SM}} -
		\frac{\Delta\Gamma_\text{total}^\text{NP}}{\Gamma_\text{total}^\text{SM}},
		\label{deltaBR}
\end{align}
where $\Delta\Gamma_i^\text{NP}$ and	$\Delta\Gamma_\text{total}^\text{NP}$
are the new physics contributions to the partial decay width $\Gamma_i^\text{SM}$
and the total decay width $\Gamma_\text{total}^\text{SM}$	 predicted by the SM, respectively.
In this equation, the first term in the right-hand side is channel dependent, while the second term is universal.
The new physics contribution to the partial width, $\Delta\Gamma_i^\text{NP}$, originates from the modifiers of the SM couplings, and the loop corrections involving the additional scalars beyond the SM (i.e. $H$, $A$, and $H^\pm$).

In Table \ref{h-BR},
we see that the most significant relative differences come from the branching fractions of the decay channels of the light Higgs boson to 
$\gamma\gamma$ and $Z\gamma$ that are 
6.2\% and 6.3\%, respectively.
These deviations are due to the radiative corrections with the charge Higgs boson propagating in the loop diagram.
Therefore, these channels can effectively distinguish the two models once the high-precision measurements are carried out in the future.
The current experimental data of the branching ratios and the total width are shown in the fifth column of Table \ref{h-BR}. 
The deviations in terms of $\sigma$ of the corresponding THDM predictions from the data, presented in the sixth column,
show that the best-fit point is in a very good agreement with the current data.
%

\section{$e^+ e^- \rightarrow t \bar{t} b \bar{b}$ channel at the ILC}
In this section, we investigate the possibility to probe the best-fit point of the type-I THDM at the ILC via the $t \bar{t} b \bar{b}$ channel.
This task has been performed using the combination of two packages,
\SARAH{} version 4.15.5
and  \GRACE{} version 2.2.1.7.

\subsection{Calculation methods}
\SARAH{} \cite{Staub:2008uz, Staub:2013tta, Staub:2009bi} is a software package written in Mathematica to investigate any models of particle physics with broken or unbroken gauge symmetries.
This tool automatically generates all possible Feynman rules for the considered model.
With the inputs of the particle content as representations of the gauge symmetry and the model's Lagrangian,
the outputs of the model are written by  \SARAH{} in the style of the CalcHEP/ComHEP\cite{Boos:1994xb, Pukhov:1999gg, Pukhov:2004ca} model files.

The \GRACE{} system is a set of programs for automatically calculating scattering amplitudes in quantum field theory developed at KEK\@.
The user can obtain numerical results of various cross-sections and decay widths by selecting the appropriate phase space option.
The public version of \GRACE{} can perform calculations in the SM and the minimal supersymmetric standard model 
\cite{Yuasa:1999rg,Fujimoto:2002sj}.
This system can be extended by adding particles and interactions to a few text files describing the model-dependent part as appropriate.
For example, an extension to the THDM and analysis using it was done in Refs.\cite{Kon:2018vmv, Grace:2006}.

\begin{table}[h!]
\begin{center}
\begin{tabular}{cc||c|c}	
\multicolumn{2}{c||}{Processes}&\multirow{2}{*}{Number of diagrams}&\multirow{2}{*}{Agreement}\\
Initial	&	Final&	&\\
\hline
$\tau^+\tau^-$&$b\bar{b}$&6&$+5.30\times10^{-14}$\\
$\tau^+\tau^-$&$t\bar{t}Z$&36& $+6.66\times10^{-16}$\\
$\nu_\mu \mu^+$&$Z W^+$&5&$-1.88\times10^{-15}$\\
$t\bar{t}$&$w^+ h^0$&5& $+8.88\times10^{-16}$\\
$t\bar{t}$&$b\bar{b}$&6& $0.00$\\
$t\bar{t}$&$h^0h^0$&6& $0.00$\\
$t\bar{t}$&$h^0h^0h^0$&88& $0.00$\\
$t\bar{t}$&$u\bar{u}$&6& $0.00$\\
$t\bar{t}$&$W^+W^-$&7& $0.00$\\
$t\bar{t}$&$Zh^0$&5& $+3.10\times10^{-15}$\\
$t\bar{t}$&$Zh^0h^0$&63& $+2.88\times10^{-15}$\\
$t\bar{t}$&$ZZ$&4& $0.00$\\
$t\bar{t}$&$h^0ZZ$&43& $+1.77\times10^{-15}$\\
$t\bar{t}$&$ZZZ$&36&$-5.55\times10^{-16}$\\
$u\bar{d}$&$h^0w^+$&5& $0.00$\\
$u\bar{d}$&$W^+ZZ$&43&$-4.99\times10^{-15}$\\
$c\bar{c}$&$W^+W^-$&7& $0.00$\\
$W^+W^-$&$\gamma\gamma$&9& $0.00$\\
$W^+W^-$&$\gamma h^0$&7&$-1.33\times10^{-15}$\\
$W^+W^-$&$h^0 h^0$&9& $+3.08\times10^{-14}$\\
$W^+W^-$&$\gamma h^0 h^0$&93& $+9.83\times10^{-12}$\\
$W^+W^-$&$W^+W^-$&9& $0.00$\\
$W^+W^-$&$Zh^0$&7&$-1.55\times10^{-15}$\\
$W^+W^-$&$Zh^0h^0$&117&$-2.72\times10^{-14}$\\
$W^+W^-$&$ZZ$&9& $0.00$\\
$ZZ$&$h^0 h^0$&9& $+1.55\times10^{-14}$\\
$ZZ$&$ZZ$&6& $0.00$	
\end{tabular}
\caption{\label{table1}\small Test for the gauge invariance}
\end{center}
\end{table}

Here, we extended \GRACE{} utilizing  \SARAH{} described above 
to perform calculations.
The \GRACE{} system has mainly two databases for generating programs to evaluate scattering amplitudes: 
a database of all particles and interaction vertices appearing in the model and a database of computer programs for numerical evaluations of a given scattering process.
The former is a list of particles with their quantum numbers, e.g., spin, parity, mass, representation in the gauge groups, etc.
All interaction vertices with corresponding interaction types are listed in the model file.
The latter is a set of computer programs to numerically evaluate spinor and vector wave functions, particle propagators with the gauge fixing parameter, interaction vertices, etc.
When the initial and final particles, as well as the interaction order (a power of coupling constants), are specified, \GRACE{} generates all possible Feynman diagrams based on the covariant gauge, including Goldstone bosons and then produces a set of computer programs to evaluate the scattering amplitude based on the generated diagrams numerically.
The amplitude is numerically integrated using an adaptive Monte Carlo method.
The \GRACE{} system also has a program database, namely the kinematics, to assign particle momenta from independent random numbers.
The \GRACE{} system keeps redundancy owing to the gauge parameters; thus, we can test generated programs if they give amplitudes independent from values of the gauge parameters.

In this study, we have generated a database of particles and interactions of the type-I THDM using \SARAH{}.
We convert outputs created by \SARAH{} to adapt with the \GRACE{} system using a Mathematica code.
The created database is numerically checked owing to the gauge-parameter independence of the system.
We have prepared 27 scattering processes concerning the fermions, the gauge bosons and the Higgs bosons covering all particles and vertices in our target process. 
We compare the numerical results with two gauge choices: the Feynman gauge and the unitary gauge. 
We obtain consistent results within the numerical precision of 64-bit floating number (binary64) format as summarized in Table \ref{table1}.

\subsection{Possible signals at the ILC}

This section discusses the discovery potential for the experimental signals of the type-I THDM with the best-fit parameters at the future linear electron-positron collider, such as the ILC.
The ILC is a linear electron-positron collider, operating at 250 to 500 GeV center-of-mass (CM) energies with high luminosity, which may be extended to 1 TeV or more in the upgrade stage. It is based on 1.3 GHz superconducting radio-frequency (SCRF) accelerating technology \cite{Behnke:2013xla, Fujii:2013lba}. 
The ILC is proposed to be constructed in the Kitakami Mountains in Tohoku, Japan. It is an international project running for more than 20 years, collaborated by more than 300 institutes, universities, and laboratories. 
The primary physics aim of the ILC is to determine the future direction of particle physics via the precise measurements of the couplings of the Higgs boson with other elementary particles. This facility can also study a wide variety of elementary-particle physics and nuclear physics. 

\begin{figure}[h!]
 \begin{center} 
   \includegraphics[width=11cm]{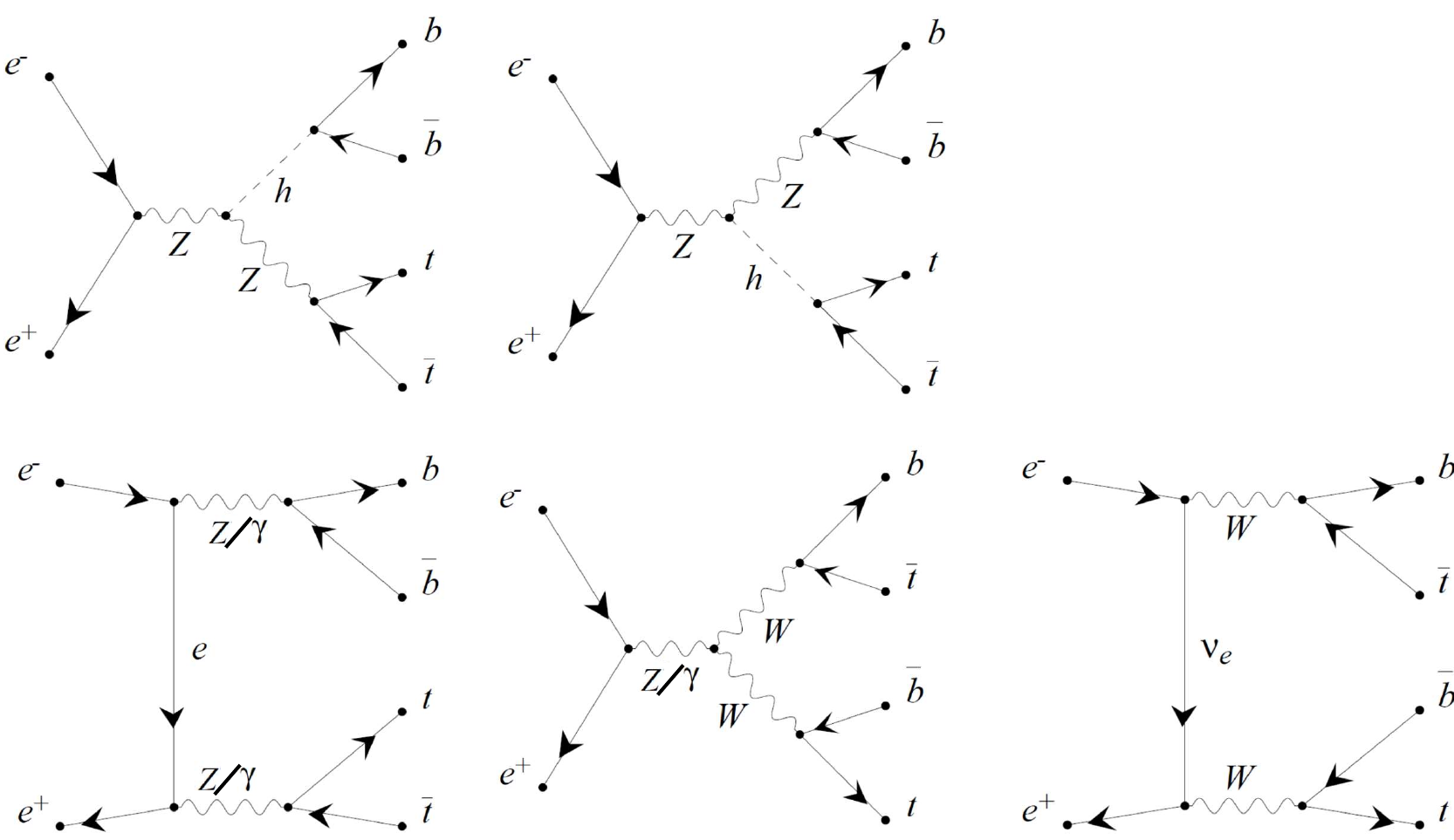}
 \caption{\label{diamsn}
Diagrams of the SM background.
The first row shows backgrounds involving the SM Higgs, while the second row shows non-Higgs backgrounds.
}
\end{center} 
\end{figure}

The prediction of the extra Higgs bosons, in addition to the standard one, is the prominent feature of the THDM.
For the experimental target, we chose the process $e^+e^-\rightarrow b\bar{b}t\bar{t}$ since the Yukawa couplings of Higgs bosons are proportional to the fermion masses, given that the $t$ and $b$ quarks are the heaviest and the second heaviest fermions, respectively.
Therefore, we can expect a sizable cross section for this process. 
This process includes pair creations of all possible Higgs bosons, such as the CP-even, CP-odd and charged Higgs particles.
Among the signal and background sub-diagrams of the target process, the light Higgs productions associated with the $Z$ boson and the gauge-boson pair-creation processes, depicted in Figure \ref{diamsn}, mainly contribute to the total cross section.
We identify the light (CP-even) Higgs boson in the THDM to the experimentally observed Higgs boson by the LHC experiments; thus, the light-Higgs production process is the background in the search for the THDM signal.
%
%
%
In our calculation,
we apply the simple condition 
$m^{~}_{b\bar{b}}>50$ GeV 
to reduce the background due to the photo-productions of the $b$-quark pair.
The main part of the SM background due to the $h$ and $Z$ resonances can be eliminated by vetoing the $b$-quark pair creations that fulfill the conditions:
\begin{align}
|M_{b\bar{b}}-m_Z| < 10~\text{GeV},~~
\text{and}~~
|M_{b\bar{b}}-m_h| < 10~\text{GeV}.
\label{veto}
\end{align}

\begin{figure}[h!]
 \begin{center} 
   \includegraphics[width=12cm]{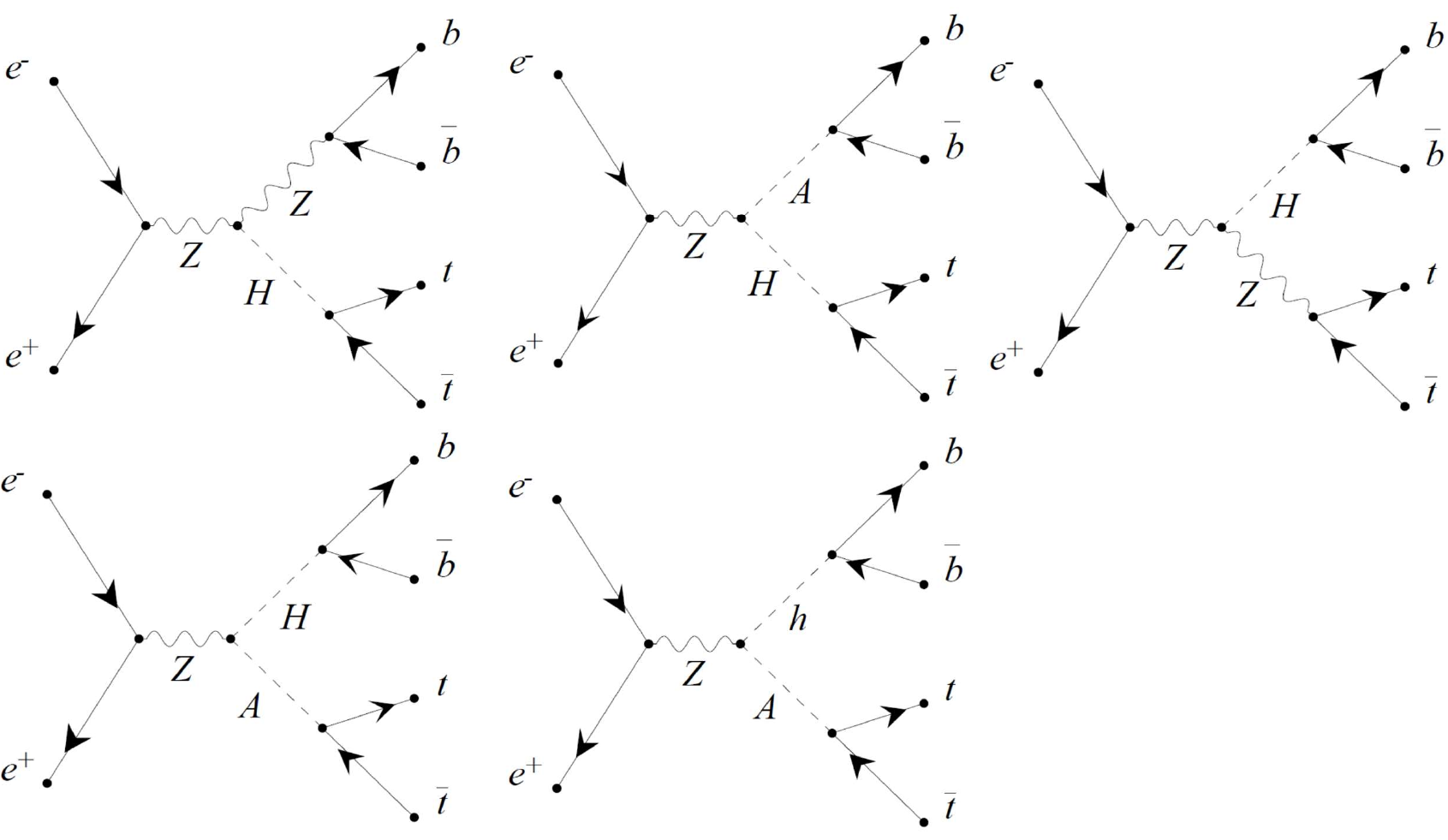}
 \caption{\label{diamnh}
Diagrams of signal processes involving the neutral Higgs bosons in the THDM.
}
\end{center} 
\end{figure}

\begin{figure}[h!]
 \begin{center} 
   \includegraphics[width=10cm]{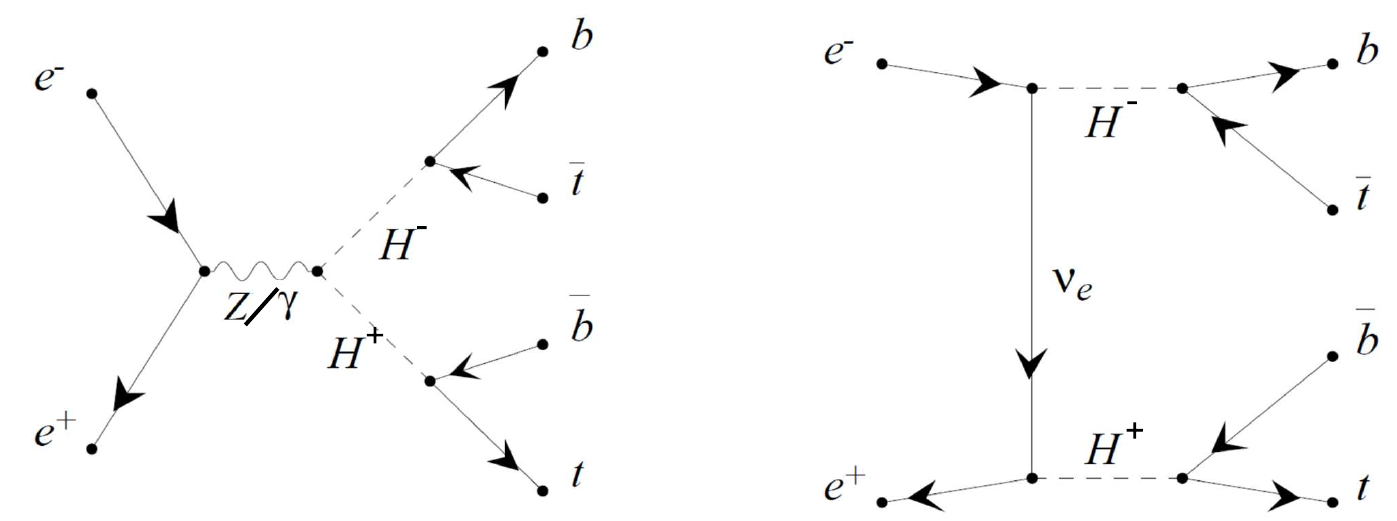}
 \caption{\label{diamch}
Diagrams of signal processes involving the charged Higgs bosons in to the THDM.
}
\end{center} 
\end{figure}

The first category of the signal processes is the pair creation of the heavy neutral Higgs bosons and the single Higgs boson production associated with a Z boson, depicted in Figure \ref{diamnh}.
The second category of the THDM signal processes consists of the charged Higgs boson pair creations depicted in Figure \ref{diamch}.
We can identify the signal processes mentioned above by looking for the resonances of $t\bar{t}$ , $b\bar{b}$ , $t\bar{b}$  and $b\bar{t}$ pairs.
All the Higgs bosons have very narrow decay widths.
Therefore, we expect the experimental signal of quark-pair invariant mass to be relatively clear after being smeared due to the detector effect.
The Feynman diagrams in Figures \ref{diamsn}-\ref{diamch} are not all the diagrams contributing to the cross section.
We have calculated the cross sections utilizing all possible Feynman diagrams (516 diagrams in total), including Goldstone bosons.
In the following calculations, we use the standard model parameters listed in Table \ref{table3} and the best-fit parameters of the THDM in Table  \ref{Bestfitpoint}.
We have calculated cross sections at the tree level using the \GRACE{} system.
The typical contributions of the higher order correction is about $\mathcal{O}(10\%)$ for the electroweak processes at the ILC energies.
It is known that the initial state photon radiation (ISR) gives the most significant contribution to the total cross sections.

\begin{table}[h!]
\begin{center}
\begin{tabular}{cc}
SM parameter & Input value \\
\hline
$m^{~}_Z$  & 91.1880 GeV		\\
$m^{~}_W$ & 80.3692 GeV	\\
$m^{~}_h$ & 125.20 GeV	\\
$\alpha(m^{~}_Z)$ &  1/128.07  \\
$\sin^2{\hspace{-.2em}\theta_W}$ & $1- {m_W^2}/{m_Z^2}$		\\
\end{tabular}
\caption{\label{table3}The SM parameters\cite{ParticleDataGroup:2024} used in calculations.}
\end{center}
\end{table}

The effect of the ISR can be factorized when the total energy of the emitted photons is sufficiently small compared to the beam energy.
The calculations under such an approximation are called the ``soft photon approximation (SPA)''.
Using the SPA, the corrected cross sections with the ISR, denoted by $\sigma^{~}_\text{ISR}$, can be obtained from the tree-level cross sections $\sigma^{~}_\text{Tree}$ using a structure function $H(x,s)$ as follows:
\begin{align} 
\sigma^{~}_\text{ISR}	&=
	\int^1_0 dx \,
	H(x,s) \,
	\sigma^{~}_\text{Tree}
	\left( s(1-x) \right),
\label{ISRTree_H}
\end{align}
where $s$ is the CM energy squared and $x$ is the energy fraction of an emitted photon.

The structure function is provided using the perturbative method with the SPA.
The radiator function $D(x,s)$, which corresponds to the square root of the structure function,  gives the probability of an electron with energy $\sqrt{s}/2$ emitting a photon with an energy fraction $x$.
In this method, electrons and positrons can emit unbalanced energies; thus, a finite boost of the CM system is realized, which is more realistic than using the structure function.
The radiator function is given as\cite{Fujimoto:1990tb}
\begin{align} 
D(1-x,s)^2&=H(x,s)=\Delta_2\beta x^{\beta-1}
-\Delta_1\beta\left(1-\frac{x}{2}\right)\nonumber\\
&+\frac{\beta^2}{8}\left[
-4(2-x)\log{x}-\frac{1+3(1-x)^2}{x}\log{(1-x)}-2x
\right],
\label{ISRLoop}
\intertext{where}
\beta	&=	\frac{2\alpha}{\pi}\left(\log{\frac{s}{m_e^2}}-1\right),
\quad ~~
\Delta_1= 1+\delta_1,~~\Delta_2 = 1+\delta_1+\delta_2,\notag\\
\delta_1 &= 
	\frac{\alpha}{\pi}
	\left(\frac{3}{2}L+\frac{\pi^2}{3}-2\right),~~~~
\delta_2 = 
	\left(\frac{\alpha L}{\pi}\right)^2
	\left(-\frac{1}{18}L+\frac{119}{72}-			\frac{\pi^2}{3}
	\right),\notag
\intertext{and}
L &= \ln \frac{s}{m_e^2}.\notag
\end{align}
This result was obtained based on the perturbative calculations for initial-state photon emission diagrams up to the two-loop order \cite{Fujimoto:1990tb}.
The terms proportional to $\alpha^2$ in Eq.(\ref{ISRLoop}) correspond to the two-loop diagrams.
The ISR corrected cross section is provided using the radiator function as
\begin{align} 
\sigma^{~}_\text{ISR}	&=
	\int^1_0 dx\int^{1}_0 dy\hspace{.1em}D(x,s)D(y,s) \,
	\sigma^{~}_\text{Tree}
	\left( sxy \right)
.
\label{ISRTree_D}
\end{align}

\begin{figure}[h!]
 \begin{center} 
   \includegraphics[width=9.5cm]{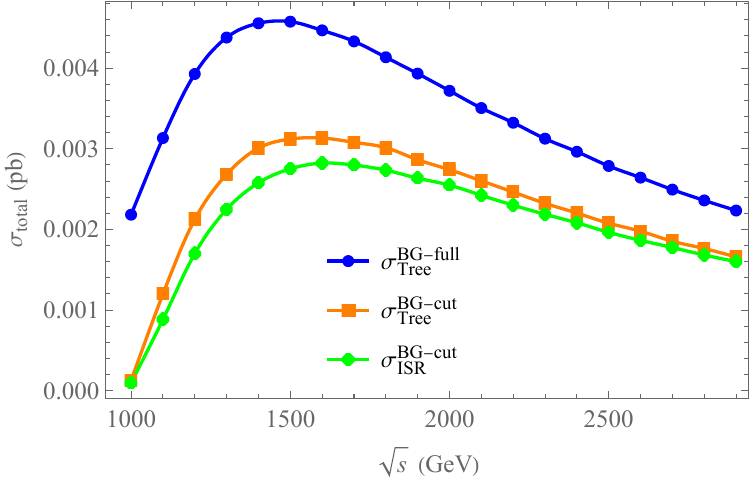}
 \caption{\label{crosssection}
The $e^+e^-\rightarrow b\bar{b}t\bar{t}$ total cross section  as a function of $\sqrt{s}$ at the ILC energies and beyond.
The blue (orange) line shows the total cross section without (with) the background cut (\ref{veto}). 
The realistic cross section including the ISR correction is shown by the green line.
}
\end{center} 
\end{figure}

\begin{figure}[h!]
 \begin{center} 
   \includegraphics[width=9.5cm]{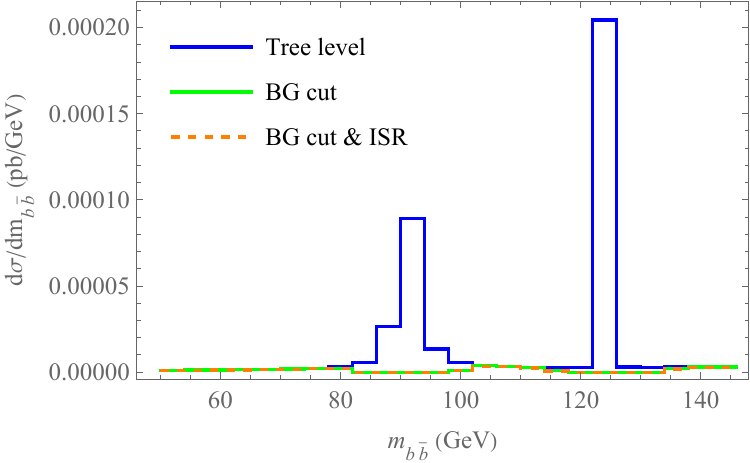}
 \caption{\label{mbb}
The $b\bar{b}$-invariant mass distributions without and with the background cut (\ref{veto}) and the ISR correction for the case $\sqrt{s} = 1500$ GeV. 
The background cut vetos events from the $Z$-boson and the light-Higgs decays.
The ISR effect is very small to be seen in the figure, leading to the overlap between the orange and the green lines.
} 
 \end{center} 
\end{figure}

\begin{figure}[h!]
 \begin{center} 
   \includegraphics[width=8cm]{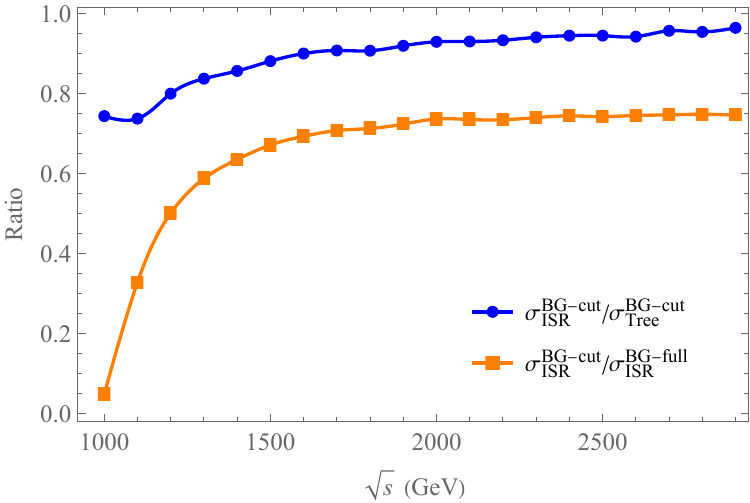}
 \caption{\label{ratio}
The reduction ratio of cross sections with the background cut and the ISR correction.
}
\end{center} 
\end{figure}

We obtain the total cross sections of the target process as shown in Figure \ref{crosssection}.
The total cross section increases from $\sqrt{s} \simeq 1000$ GeV owing to the charged-Higgs pair-creation threshold, 
peaks around 1500 GeV, 
and then gradually decreases according to the increase in the CM energy.
Among all diagrams of the target process, the SM-like Higgs production associated with the Z boson gives the most significant background, as shown in Figure \ref{mbb}.
The figure shows the clear twin peaks of the $b$-quark pair invariant mass at the $Z$-boson and the light-Higgs masses.
After cutting the events satisfying (\ref{veto}) around the $Z$ and Higgs masses, the charged Higgs pair production cross section is dominated, while other Higgs contributions (the heavy and CP-odd Higgs) are negligibly small.
The cuts (\ref{veto}) eliminate about $30\%$ of the total events in high energy region as shown by the orange line in Figure \ref{ratio}, which is 
consistent to the SM background cross section.
The ISR correction leads to the effective CM energy smaller than the nominal energy due to the lost in term of radiation; thus, the signal cross sections are decreasing about $10$ to $20$\% of the total cross sections after applying this correction, also shown by the blue line in Figure \ref{ratio}.

\begin{figure}[h!]
 \begin{center} 
   \includegraphics[width=9.5cm]{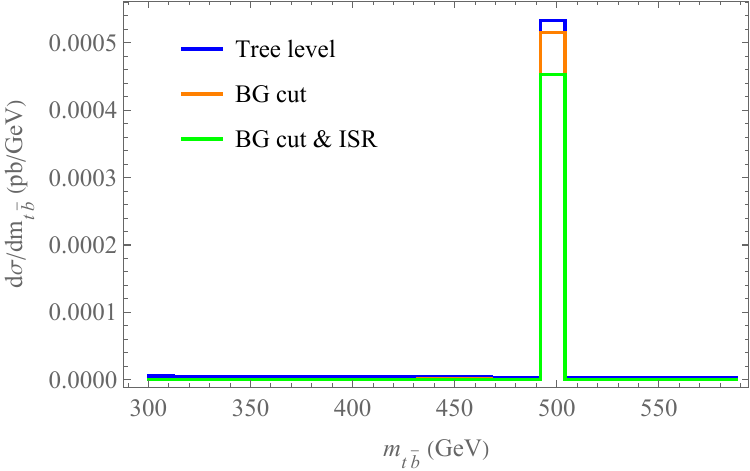}
 \caption{\label{tb}
The $t\bar{b}\,(b\bar{t})$-invariant mass distributions without and with the background cut and the ISR correction for the case $\sqrt{s} = 1500$ GeV. 
The background cut does not significantly reduce the number of signal events.
} 
 \end{center} 
\end{figure}

\begin{figure}[h!]
 \begin{center} 
   \includegraphics[width=9.5cm]{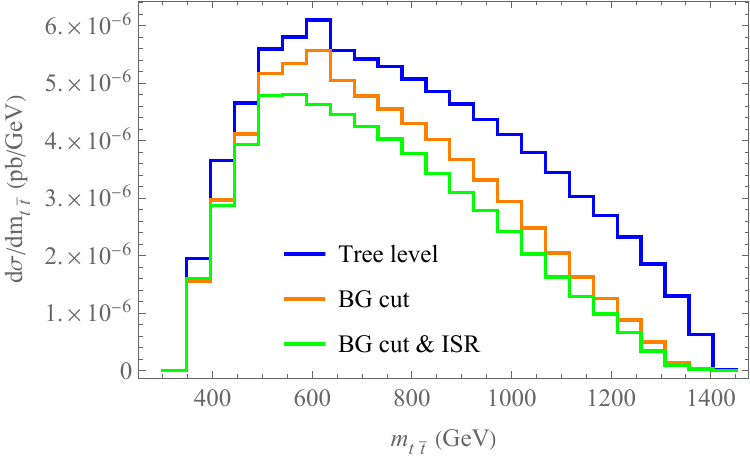}
 \caption{\label{mtt}
The $t\bar{t}$-invariant mass distributions without and with the background cut and the ISR correction for the case $\sqrt{s} = 1500$ GeV. 
} 
 \end{center} 
\end{figure}

The charged Higgs signal is obvious in the $t\bar{b}$ ($b\bar{t}$) invariant mass distribution, as shown in Figure \ref{tb}.
The signal peak is significant even before the background cut because the SM does not have any structure on the $t\bar{b}$ invariant mass distribution, and the phase space contribution in this energy region is very small.
The ISR affects the mass distribution only $10$\% effect; thus, we can expect a clear signal of the charged Higgs production after the smearing due to the detector effects. 
%
%
On the other hand, the $t$-quark pair invariant mass does not have any structure, as shown in Figure~\ref{mtt}.
When the heavy CP-even and CP-odd Higgs production, depicted in Figure \ref{diamnh}, have significant cross sections, we can observe peak(s) on this distribution.
However, the THDM with the best-fit parameter set gives a small differential cross section with respect to $m_{t\bar{t}}$.
Thus, it is very challenging to experimentally observe the signal via the $m_{t\bar{t}}$ distribution.

After the experimental cuts and the ISR correction, the total cross sections are greater than $1$fb for the CM energy greater than 1100 GeV.
Therefore, if we assume $10$ ab$^{-1}$  of an integrated luminosity at the ILC, we can expect $\mathcal{O}(10^{4})$ almost-background-free signal events.
Even if we assume several per cent of the reconstruction efficiency of the $t$ and $b$ quarks, we can easily distinguish the THDM signal from the SM background.

\section{Summary}
We have investigated the phenomenology of the type-I THDM, an extension of the SM adding one more Higgs doublet that does not couple directly to fermions via the Yukawa interaction.
The Lagrangian satisfies the $Z_2$ symmetry except for the softly breaking term in the scalar potential.
We took into account various theoretical and phenomenological constraints such as the perturbativity, unitarity, stability, oblique parameters, the Higgs searches by high-energy experiments (LEP, Tevatron and LHC ), the world-average $W$-boson mass, as well as the flavour physics observables.
We found the best-fit point to the measured observables by performing a systematic scan over the space of independent parameters.
The pulls for the branching ratios of the decays 
$D_s^+ \rightarrow \tau^+ \nu_\tau$, $D^+ \rightarrow \mu^+ \nu_\mu$, and $B^0 \rightarrow K^* \mu^+ \mu^-$ at low momentum transfer are noticeable.
Therefore, they are suitable probes to test the viability of the best-fit point by more precise measurements of these channels in the future.
The branching ratios of the SM-like Higgs boson decaying to 
$\gamma\gamma$ and $Z\gamma$ deviates from the SM value at the level of 
6.2\% and 6.3\%, respectively.
The results 
motivate more precise analyses of these light-Higgs decay channels at future colliders to discriminate between the two models.

Considering the possibility of observing the THDM new physics at one of the most promising future colliders, the ILC, we have proposed the analysis of the best-fit point using the $e^+e^- \rightarrow t\bar{t} b\bar{b}$ scattering process.
We have utilized the \GRACE{} system to calculate the total cross-sections and the invariant mass distributions.
Since the standard \GRACE{} system does not have a model file of the THDM, we have prepared it automatically using the \SARAH{} system.
We have tested the consistency of the created model file utilizing the gauge-parameter independence of the scattering amplitudes for various scattering processes.
We have considered the ISR effect in the calculation, which is known to contribute most to the total cross-sections at the one-loop level. 
Consequently, we clarified that the THDM with the best-fit parameter set can be probed by observing the clear charged Higgs signal on the $t\bar{b}$ invariant mass distribution.
On the other hand, the signals of other Higgs bosons included in the THDM, the heavy CP-even Higgs boson and the CP-odd Higgs boson, are hardly observed at the ILC since they do not provide significant cross sections.
The reasons are that the best-fit point predicts the light Higgs boson consistent with observations to date, leading to the small couplings of other heavy neutral Higgs bosons to the fermions.
Meanwhile, the couplings of charged Higgs to the heavy fermions are of $\mathcal{O}(1)$.

\section*{Acknowledgements}

Hieu Minh Tran is supported by Vietnam 
National Foundation for Science and Technology Development (NAFOSTED)
under grant number 103.01-2023.75.
Huong Thu Nguyen is supported by the research project QG.21.15 of Vietnam National University, Hanoi.


\bibliographystyle{JHEP} 
\bibliography{references}

\providecommand{\href}[2]{#2}\begingroup\raggedright\begin{thebibliography}{100}

\bibitem{Haber:1978jt}
H.E.~Haber, G.L.~Kane and T.~Sterling,  \textit{{The Fermion Mass Scale and
  Possible Effects of Higgs Bosons on Experimental Observables}},
  \href{https://doi.org/10.1016/0550-3213(79)90225-6}{Nucl. Phys. B {\bfseries
  161} (1979) 493}.

\bibitem{Martin:1997ns}
S.P.~Martin,  \textit{{A Supersymmetry primer}},
  \href{https://doi.org/10.1142/9789812839657_0001}{Adv. Ser. Direct. High
  Energy Phys. {\bfseries 18} (1998) 1}
  [\href{https://arxiv.org/abs/hep-ph/9709356}{{\ttfamily hep-ph/9709356}}].

\bibitem{Okada:2010xe}
N.~Okada and H.M.~Tran,  \textit{{Discrimination of Supersymmetric Grand
  Unified Models in Gaugino Mediation}},
  \href{https://doi.org/10.1103/PhysRevD.83.053001}{Phys. Rev. D {\bfseries 83}
  (2011) 053001} [\href{https://arxiv.org/abs/1011.1668}{{\ttfamily
  1011.1668}}].

\bibitem{Tran:2010ea}
H.M.~Tran, T.~Kon and Y.~Kurihara,  \textit{{Discrimination of SUSY breaking
  models using single-photon processes at future e+e- linear colliders}},
  \href{https://doi.org/10.1142/S0217732311035420}{Mod. Phys. Lett. A
  {\bfseries 26} (2011) 949} [\href{https://arxiv.org/abs/1012.1730}{{\ttfamily
  1012.1730}}].

\bibitem{Croon:2019kpe}
D.~Croon, T.E.~Gonzalo, L.~Graf, N.~Ko\v{s}nik and G.~White,  \textit{{GUT
  Physics in the era of the LHC}},
  \href{https://doi.org/10.3389/fphy.2019.00076}{Front. in Phys. {\bfseries 7}
  (2019) 76} [\href{https://arxiv.org/abs/1903.04977}{{\ttfamily 1903.04977}}].

\bibitem{Fukuyama:2019zun}
T.~Fukuyama, N.~Okada and H.M.~Tran,  \textit{{Alternative renormalizable
  $SO$(10) GUTs and data fitting}},
  \href{https://doi.org/10.1016/j.nuclphysb.2020.114992}{Nucl. Phys. B
  {\bfseries 954} (2020) 114992}
  [\href{https://arxiv.org/abs/1907.02948}{{\ttfamily 1907.02948}}].

\bibitem{Peccei:1977hh}
R.D.~Peccei and H.R.~Quinn,  \textit{{CP Conservation in the Presence of
  Instantons}}, \href{https://doi.org/10.1103/PhysRevLett.38.1440}{Phys. Rev.
  Lett. {\bfseries 38} (1977) 1440}.

\bibitem{Peccei:1977ur}
R.D.~Peccei and H.R.~Quinn,  \textit{{Constraints Imposed by CP Conservation in
  the Presence of Instantons}},
  \href{https://doi.org/10.1103/PhysRevD.16.1791}{Phys. Rev. D {\bfseries 16}
  (1977) 1791}.

\bibitem{Turok:1990in}
N.~Turok and J.~Zadrozny,  \textit{{Dynamical generation of baryons at the
  electroweak transition}},
  \href{https://doi.org/10.1103/PhysRevLett.65.2331}{Phys. Rev. Lett.
  {\bfseries 65} (1990) 2331}.

\bibitem{Nelson:1991ab}
A.E.~Nelson, D.B.~Kaplan and A.G.~Cohen,  \textit{{Why there is something
  rather than nothing: Matter from weak interactions}},
  \href{https://doi.org/10.1016/0550-3213(92)90440-M}{Nucl. Phys. B {\bfseries
  373} (1992) 453}.

\bibitem{Funakubo:1993jg}
K.~Funakubo, A.~Kakuto and K.~Takenaga,  \textit{{The Effective potential of
  electroweak theory with two massless Higgs doublets at finite temperature}},
  \href{https://doi.org/10.1143/PTP.91.341}{Prog. Theor. Phys. {\bfseries 91}
  (1994) 341} [\href{https://arxiv.org/abs/hep-ph/9310267}{{\ttfamily
  hep-ph/9310267}}].

\bibitem{Cline:1995dg}
J.M.~Cline, K.~Kainulainen and A.P.~Vischer,  \textit{{Dynamics of two Higgs
  doublet CP violation and baryogenesis at the electroweak phase transition}},
  \href{https://doi.org/10.1103/PhysRevD.54.2451}{Phys. Rev. D {\bfseries 54}
  (1996) 2451} [\href{https://arxiv.org/abs/hep-ph/9506284}{{\ttfamily
  hep-ph/9506284}}].

\bibitem{Cline:1996mga}
J.M.~Cline and P.-A.~Lemieux,  \textit{{Electroweak phase transition in two
  Higgs doublet models}}, \href{https://doi.org/10.1103/PhysRevD.55.3873}{Phys.
  Rev. D {\bfseries 55} (1997) 3873}
  [\href{https://arxiv.org/abs/hep-ph/9609240}{{\ttfamily hep-ph/9609240}}].

\bibitem{Laine:2000rm}
M.~Laine and K.~Rummukainen,  \textit{{Two Higgs doublet dynamics at the
  electroweak phase transition: A Nonperturbative study}},
  \href{https://doi.org/10.1016/S0550-3213(00)00736-7}{Nucl. Phys. B {\bfseries
  597} (2001) 23} [\href{https://arxiv.org/abs/hep-lat/0009025}{{\ttfamily
  hep-lat/0009025}}].

\bibitem{Fromme:2006cm}
L.~Fromme, S.J.~Huber and M.~Seniuch,  \textit{{Baryogenesis in the two-Higgs
  doublet model}}, \href{https://doi.org/10.1088/1126-6708/2006/11/038}{JHEP
  {\bfseries 11} (2006) 038}
  [\href{https://arxiv.org/abs/hep-ph/0605242}{{\ttfamily hep-ph/0605242}}].

\bibitem{Davies:1994id}
A.T.~Davies, C.D.~froggatt, G.~Jenkins and R.G.~Moorhouse,
  \textit{{Baryogenesis constraints on two Higgs doublet models}},
  \href{https://doi.org/10.1016/0370-2693(94)90559-2}{Phys. Lett. B {\bfseries
  336} (1994) 464}.

\bibitem{Branco:2011iw}
G.C.~Branco, P.M.~Ferreira, L.~Lavoura, M.N.~Rebelo, M.~Sher and J.P.~Silva,
  \textit{{Theory and phenomenology of two-Higgs-doublet models}},
  \href{https://doi.org/10.1016/j.physrep.2012.02.002}{Phys. Rept. {\bfseries
  516} (2012) 1} [\href{https://arxiv.org/abs/1106.0034}{{\ttfamily
  1106.0034}}].

\bibitem{Glashow:1976nt}
S.L.~Glashow and S.~Weinberg,  \textit{{Natural Conservation Laws for Neutral
  Currents}}, \href{https://doi.org/10.1103/PhysRevD.15.1958}{Phys. Rev. D
  {\bfseries 15} (1977) 1958}.

\bibitem{Manohar:2006ga}
A.V.~Manohar and M.B.~Wise,  \textit{{Flavor changing neutral currents, an
  extended scalar sector, and the Higgs production rate at the CERN LHC}},
  \href{https://doi.org/10.1103/PhysRevD.74.035009}{Phys. Rev. D {\bfseries 74}
  (2006) 035009} [\href{https://arxiv.org/abs/hep-ph/0606172}{{\ttfamily
  hep-ph/0606172}}].

\bibitem{Pich:2009sp}
A.~Pich and P.~Tuzon,  \textit{{Yukawa Alignment in the Two-Higgs-Doublet
  Model}}, \href{https://doi.org/10.1103/PhysRevD.80.091702}{Phys. Rev. D
  {\bfseries 80} (2009) 091702}
  [\href{https://arxiv.org/abs/0908.1554}{{\ttfamily 0908.1554}}].

\bibitem{Penuelas:2017ikk}
A.~Pe\~nuelas and A.~Pich,  \textit{{Flavour alignment in multi-Higgs-doublet
  models}}, \href{https://doi.org/10.1007/JHEP12(2017)084}{JHEP {\bfseries 12}
  (2017) 084} [\href{https://arxiv.org/abs/1710.02040}{{\ttfamily
  1710.02040}}].

\bibitem{Jung:2010ik}
M.~Jung, A.~Pich and P.~Tuzon,  \textit{{Charged-Higgs phenomenology in the
  Aligned two-Higgs-doublet model}},
  \href{https://doi.org/10.1007/JHEP11(2010)003}{JHEP {\bfseries 11} (2010)
  003} [\href{https://arxiv.org/abs/1006.0470}{{\ttfamily 1006.0470}}].

\bibitem{Chen:2013kt}
C.-Y.~Chen and S.~Dawson,  \textit{{Exploring Two Higgs Doublet Models Through
  Higgs Production}}, \href{https://doi.org/10.1103/PhysRevD.87.055016}{Phys.
  Rev. D {\bfseries 87} (2013) 055016}
  [\href{https://arxiv.org/abs/1301.0309}{{\ttfamily 1301.0309}}].

\bibitem{Celis:2013rcs}
A.~Celis, V.~Ilisie and A.~Pich,  \textit{{LHC constraints on two-Higgs doublet
  models}}, \href{https://doi.org/10.1007/JHEP07(2013)053}{JHEP {\bfseries 07}
  (2013) 053} [\href{https://arxiv.org/abs/1302.4022}{{\ttfamily 1302.4022}}].

\bibitem{Chiang:2013ixa}
C.-W.~Chiang and K.~Yagyu,  \textit{{Implications of Higgs boson search data on
  the two-Higgs doublet models with a softly broken $Z_2$ symmetry}},
  \href{https://doi.org/10.1007/JHEP07(2013)160}{JHEP {\bfseries 07} (2013)
  160} [\href{https://arxiv.org/abs/1303.0168}{{\ttfamily 1303.0168}}].

\bibitem{Grinstein:2013npa}
B.~Grinstein and P.~Uttayarat,  \textit{{Carving Out Parameter Space in Type-II
  Two Higgs Doublets Model}},
  \href{https://doi.org/10.1007/JHEP06(2013)094}{JHEP {\bfseries 06} (2013)
  094} [\href{https://arxiv.org/abs/1304.0028}{{\ttfamily 1304.0028}}].

\bibitem{Eberhardt:2013uba}
O.~Eberhardt, U.~Nierste and M.~Wiebusch,  \textit{{Status of the
  two-Higgs-doublet model of type II}},
  \href{https://doi.org/10.1007/JHEP07(2013)118}{JHEP {\bfseries 07} (2013)
  118} [\href{https://arxiv.org/abs/1305.1649}{{\ttfamily 1305.1649}}].

\bibitem{Celis:2013ixa}
A.~Celis, V.~Ilisie and A.~Pich,  \textit{{Towards a general analysis of LHC
  data within two-Higgs-doublet models}},
  \href{https://doi.org/10.1007/JHEP12(2013)095}{JHEP {\bfseries 12} (2013)
  095} [\href{https://arxiv.org/abs/1310.7941}{{\ttfamily 1310.7941}}].

\bibitem{Chang:2013ona}
S.~Chang, S.K.~Kang, J.-P.~Lee, K.Y.~Lee, S.C.~Park and J.~Song,  \textit{{Two
  Higgs doublet models for the LHC Higgs boson data at $\sqrt{s}=$ 7 and 8
  TeV}}, \href{https://doi.org/10.1007/JHEP09(2014)101}{JHEP {\bfseries 09}
  (2014) 101} [\href{https://arxiv.org/abs/1310.3374}{{\ttfamily 1310.3374}}].

\bibitem{Wang:2013sha}
L.~Wang and X.-F.~Han,  \textit{{Status of the aligned two-Higgs-doublet model
  confronted with the Higgs data}},
  \href{https://doi.org/10.1007/JHEP04(2014)128}{JHEP {\bfseries 04} (2014)
  128} [\href{https://arxiv.org/abs/1312.4759}{{\ttfamily 1312.4759}}].

\bibitem{Baglio:2014nea}
J.~Baglio, O.~Eberhardt, U.~Nierste and M.~Wiebusch,  \textit{{Benchmarks for
  Higgs Pair Production and Heavy Higgs boson Searches in the Two-Higgs-Doublet
  Model of Type II}}, \href{https://doi.org/10.1103/PhysRevD.90.015008}{Phys.
  Rev. D {\bfseries 90} (2014) 015008}
  [\href{https://arxiv.org/abs/1403.1264}{{\ttfamily 1403.1264}}].

\bibitem{Ilisie:2014hea}
V.~Ilisie and A.~Pich,  \textit{{Low-mass fermiophobic charged Higgs
  phenomenology in two-Higgs-doublet models}},
  \href{https://doi.org/10.1007/JHEP09(2014)089}{JHEP {\bfseries 09} (2014)
  089} [\href{https://arxiv.org/abs/1405.6639}{{\ttfamily 1405.6639}}].

\bibitem{Kanemura:2014bqa}
S.~Kanemura, K.~Tsumura, K.~Yagyu and H.~Yokoya,  \textit{{Fingerprinting
  nonminimal Higgs sectors}},
  \href{https://doi.org/10.1103/PhysRevD.90.075001}{Phys. Rev. D {\bfseries 90}
  (2014) 075001} [\href{https://arxiv.org/abs/1406.3294}{{\ttfamily
  1406.3294}}].

\bibitem{Bernon:2014nxa}
J.~Bernon, J.F.~Gunion, Y.~Jiang and S.~Kraml,  \textit{{Light Higgs bosons in
  Two-Higgs-Doublet Models}},
  \href{https://doi.org/10.1103/PhysRevD.91.075019}{Phys. Rev. D {\bfseries 91}
  (2015) 075019} [\href{https://arxiv.org/abs/1412.3385}{{\ttfamily
  1412.3385}}].

\bibitem{Craig:2015jba}
N.~Craig, F.~D'Eramo, P.~Draper, S.~Thomas and H.~Zhang,  \textit{{The Hunt for
  the Rest of the Higgs Bosons}},
  \href{https://doi.org/10.1007/JHEP06(2015)137}{JHEP {\bfseries 06} (2015)
  137} [\href{https://arxiv.org/abs/1504.04630}{{\ttfamily 1504.04630}}].

\bibitem{Abbas:2015cua}
G.~Abbas, A.~Celis, X.-Q.~Li, J.~Lu and A.~Pich,  \textit{{Flavour-changing top
  decays in the aligned two-Higgs-doublet model}},
  \href{https://doi.org/10.1007/JHEP06(2015)005}{JHEP {\bfseries 06} (2015)
  005} [\href{https://arxiv.org/abs/1503.06423}{{\ttfamily 1503.06423}}].

\bibitem{Botella:2015hoa}
F.J.~Botella, G.C.~Branco, M.~Nebot and M.N.~Rebelo,  \textit{{Flavour Changing
  Higgs Couplings in a Class of Two Higgs Doublet Models}},
  \href{https://doi.org/10.1140/epjc/s10052-016-3993-0}{Eur. Phys. J. C
  {\bfseries 76} (2016) 161}
  [\href{https://arxiv.org/abs/1508.05101}{{\ttfamily 1508.05101}}].

\bibitem{Ilnicka:2015jba}
A.~Ilnicka, M.~Krawczyk and T.~Robens,  \textit{{Inert Doublet Model in light
  of LHC Run I and astrophysical data}},
  \href{https://doi.org/10.1103/PhysRevD.93.055026}{Phys. Rev. D {\bfseries 93}
  (2016) 055026} [\href{https://arxiv.org/abs/1508.01671}{{\ttfamily
  1508.01671}}].

\bibitem{Bernon:2015qea}
J.~Bernon, J.F.~Gunion, H.E.~Haber, Y.~Jiang and S.~Kraml,
  \textit{{Scrutinizing the alignment limit in two-Higgs-doublet models:
  m$_h$=125 GeV}}, \href{https://doi.org/10.1103/PhysRevD.92.075004}{Phys. Rev.
  D {\bfseries 92} (2015) 075004}
  [\href{https://arxiv.org/abs/1507.00933}{{\ttfamily 1507.00933}}].

\bibitem{Bernon:2015wef}
J.~Bernon, J.F.~Gunion, H.E.~Haber, Y.~Jiang and S.~Kraml,
  \textit{{Scrutinizing the alignment limit in two-Higgs-doublet models. II.
  m$_H$=125 GeV}}, \href{https://doi.org/10.1103/PhysRevD.93.035027}{Phys. Rev.
  D {\bfseries 93} (2016) 035027}
  [\href{https://arxiv.org/abs/1511.03682}{{\ttfamily 1511.03682}}].

\bibitem{Cacchio:2016qyh}
V.~Cacchio, D.~Chowdhury, O.~Eberhardt and C.W.~Murphy,
  \textit{{Next-to-leading order unitarity fits in Two-Higgs-Doublet models
  with soft $\mathbb{Z}_2$ breaking}},
  \href{https://doi.org/10.1007/JHEP11(2016)026}{JHEP {\bfseries 11} (2016)
  026} [\href{https://arxiv.org/abs/1609.01290}{{\ttfamily 1609.01290}}].

\bibitem{Belusca-Maito:2016dqe}
H.~B\'elusca-Ma\"\i{}to, A.~Falkowski, D.~Fontes, J.C.~Rom\~ao and
  J.a.P.~Silva,  \textit{{Higgs EFT for 2HDM and beyond}},
  \href{https://doi.org/10.1140/epjc/s10052-017-4745-5}{Eur. Phys. J. C
  {\bfseries 77} (2017) 176}
  [\href{https://arxiv.org/abs/1611.01112}{{\ttfamily 1611.01112}}].

\bibitem{Ilnicka:2018def}
A.~Ilnicka, T.~Robens and T.~Stefaniak,  \textit{{Constraining Extended Scalar
  Sectors at the LHC and beyond}},
  \href{https://doi.org/10.1142/S0217732318300070}{Mod. Phys. Lett. A
  {\bfseries 33} (2018) 1830007}
  [\href{https://arxiv.org/abs/1803.03594}{{\ttfamily 1803.03594}}].

\bibitem{Dercks:2018wch}
D.~Dercks and T.~Robens,  \textit{{Constraining the Inert Doublet Model using
  Vector Boson Fusion}},
  \href{https://doi.org/10.1140/epjc/s10052-019-7436-6}{Eur. Phys. J. C
  {\bfseries 79} (2019) 924}
  [\href{https://arxiv.org/abs/1812.07913}{{\ttfamily 1812.07913}}].

\bibitem{Botella:2018gzy}
F.J.~Botella, F.~Cornet-Gomez and M.~Nebot,  \textit{{Flavor conservation in
  two-Higgs-doublet models}},
  \href{https://doi.org/10.1103/PhysRevD.98.035046}{Phys. Rev. D {\bfseries 98}
  (2018) 035046} [\href{https://arxiv.org/abs/1803.08521}{{\ttfamily
  1803.08521}}].

\bibitem{Sanyal:2019xcp}
P.~Sanyal,  \textit{{Limits on the Charged Higgs Parameters in the Two Higgs
  Doublet Model using CMS $\sqrt{s}=13$ TeV Results}},
  \href{https://doi.org/10.1140/epjc/s10052-019-7431-y}{Eur. Phys. J. C
  {\bfseries 79} (2019) 913}
  [\href{https://arxiv.org/abs/1906.02520}{{\ttfamily 1906.02520}}].

\bibitem{Herrero-Garcia:2019mcy}
J.~Herrero-Garcia, M.~Nebot, F.~Rajec, M.~White and A.G.~Williams,
  \textit{{Higgs Quark Flavor Violation: Simplified Models and Status of
  General Two-Higgs-Doublet Model}},
  \href{https://doi.org/10.1007/JHEP02(2020)147}{JHEP {\bfseries 02} (2020)
  147} [\href{https://arxiv.org/abs/1907.05900}{{\ttfamily 1907.05900}}].

\bibitem{Karmakar:2019vnq}
S.~Karmakar and S.~Rakshit,  \textit{{Relaxed constraints on the heavy scalar
  masses in 2HDM}}, \href{https://doi.org/10.1103/PhysRevD.100.055016}{Phys.
  Rev. D {\bfseries 100} (2019) 055016}
  [\href{https://arxiv.org/abs/1901.11361}{{\ttfamily 1901.11361}}].

\bibitem{Chen:2019pkq}
N.~Chen, T.~Han, S.~Li, S.~Su, W.~Su and Y.~Wu,  \textit{{Type-I 2HDM under the
  Higgs and Electroweak Precision Measurements}},
  \href{https://doi.org/10.1007/JHEP08(2020)131}{JHEP {\bfseries 08} (2020)
  131} [\href{https://arxiv.org/abs/1912.01431}{{\ttfamily 1912.01431}}].

\bibitem{Arco:2020ucn}
F.~Arco, S.~Heinemeyer and M.J.~Herrero,  \textit{{Exploring sizable triple
  Higgs couplings in the 2HDM}},
  \href{https://doi.org/10.1140/epjc/s10052-020-8406-8}{Eur. Phys. J. C
  {\bfseries 80} (2020) 884}
  [\href{https://arxiv.org/abs/2005.10576}{{\ttfamily 2005.10576}}].

\bibitem{Aiko:2020ksl}
M.~Aiko, S.~Kanemura, M.~Kikuchi, K.~Mawatari, K.~Sakurai and K.~Yagyu,
  \textit{{Probing extended Higgs sectors by the synergy between direct
  searches at the LHC and precision tests at future lepton colliders}},
  \href{https://doi.org/10.1016/j.nuclphysb.2021.115375}{Nucl. Phys. B
  {\bfseries 966} (2021) 115375}
  [\href{https://arxiv.org/abs/2010.15057}{{\ttfamily 2010.15057}}].

\bibitem{Botella:2020xzf}
F.J.~Botella, F.~Cornet-Gomez and M.~Nebot,  \textit{{Electron and muon $g-2$
  anomalies in general flavour conserving two Higgs doublets models}},
  \href{https://doi.org/10.1103/PhysRevD.102.035023}{Phys. Rev. D {\bfseries
  102} (2020) 035023} [\href{https://arxiv.org/abs/2006.01934}{{\ttfamily
  2006.01934}}].

\bibitem{Athron:2021auq}
P.~Athron, C.~Balazs, T.E.~Gonzalo, D.~Jacob, F.~Mahmoudi and C.~Sierra,
  \textit{{Likelihood analysis of the flavour anomalies and g \textendash{} 2
  in the general two Higgs doublet model}},
  \href{https://doi.org/10.1007/JHEP01(2022)037}{JHEP {\bfseries 01} (2022)
  037} [\href{https://arxiv.org/abs/2111.10464}{{\ttfamily 2111.10464}}].

\bibitem{Botella:2022rte}
F.J.~Botella, F.~Cornet-Gomez, C.~Mir\'o and M.~Nebot,  \textit{{Muon and
  electron $g-2$ anomalies in a flavor conserving 2HDM with an oblique view on
  the CDF $M_W$ value}},
  \href{https://doi.org/10.1140/epjc/s10052-022-10893-x}{Eur. Phys. J. C
  {\bfseries 82} (2022) 915}
  [\href{https://arxiv.org/abs/2205.01115}{{\ttfamily 2205.01115}}].

\bibitem{Botella:2023tiw}
F.J.~Botella, F.~Cornet-Gomez, C.~Mir\'o and M.~Nebot,  \textit{{New Physics
  hints from $\tau$ scalar interactions and $(g-2)_{e,\mu}$}},
  \href{https://arxiv.org/abs/2302.05471}{{\ttfamily 2302.05471}}.

\bibitem{Chowdhury:2015yja}
D.~Chowdhury and O.~Eberhardt,  \textit{{Global fits of the two-loop
  renormalized Two-Higgs-Doublet model with soft Z$_{2}$ breaking}},
  \href{https://doi.org/10.1007/JHEP11(2015)052}{JHEP {\bfseries 11} (2015)
  052} [\href{https://arxiv.org/abs/1503.08216}{{\ttfamily 1503.08216}}].

\bibitem{Chowdhury:2017aav}
D.~Chowdhury and O.~Eberhardt,  \textit{{Update of Global Two-Higgs-Doublet
  Model Fits}}, \href{https://doi.org/10.1007/JHEP05(2018)161}{JHEP {\bfseries
  05} (2018) 161} [\href{https://arxiv.org/abs/1711.02095}{{\ttfamily
  1711.02095}}].

\bibitem{Haller:2018nnx}
J.~Haller, A.~Hoecker, R.~Kogler, K.~M\"onig, T.~Peiffer and J.~Stelzer,
  \textit{{Update of the global electroweak fit and constraints on
  two-Higgs-doublet models}},
  \href{https://doi.org/10.1140/epjc/s10052-018-6131-3}{Eur. Phys. J. C
  {\bfseries 78} (2018) 675}
  [\href{https://arxiv.org/abs/1803.01853}{{\ttfamily 1803.01853}}].

\bibitem{Eberhardt:2020dat}
O.~Eberhardt, A.P.~Mart\'\i{}nez and A.~Pich,  \textit{{Global fits in the
  Aligned Two-Higgs-Doublet model}},
  \href{https://doi.org/10.1007/JHEP05(2021)005}{JHEP {\bfseries 05} (2021)
  005} [\href{https://arxiv.org/abs/2012.09200}{{\ttfamily 2012.09200}}].

\bibitem{Karan:2023kyj}
A.~Karan, V.~Miralles and A.~Pich,  \textit{{Updated global fit of the ATHDM
  with heavy scalars}},  \href{https://arxiv.org/abs/2307.15419}{{\ttfamily
  2307.15419}}.

\bibitem{Kling:2018xud}
F.~Kling, H.~Li, A.~Pyarelal, H.~Song and S.~Su,  \textit{{Exotic Higgs Decays
  in Type-II 2HDMs at the LHC and Future 100 TeV Hadron Colliders}},
  \href{https://doi.org/10.1007/JHEP06(2019)031}{JHEP {\bfseries 06} (2019)
  031} [\href{https://arxiv.org/abs/1812.01633}{{\ttfamily 1812.01633}}].

\bibitem{Adhikary:2018ise}
A.~Adhikary, S.~Banerjee, R.~Kumar~Barman and B.~Bhattacherjee,
  \textit{{Resonant heavy Higgs searches at the HL-LHC}},
  \href{https://doi.org/10.1007/JHEP09(2019)068}{JHEP {\bfseries 09} (2019)
  068} [\href{https://arxiv.org/abs/1812.05640}{{\ttfamily 1812.05640}}].

\bibitem{Kon:2018vmv}
T.~Kon, T.~Nagura, T.~Ueda and K.~Yagyu,  \textit{{Double Higgs boson
  production at $e^+e^-$ colliders in the two-Higgs-doublet model}},
  \href{https://doi.org/10.1103/PhysRevD.99.095027}{Phys. Rev. D {\bfseries 99}
  (2019) 095027} [\href{https://arxiv.org/abs/1812.09843}{{\ttfamily
  1812.09843}}].

\bibitem{Arco:2021bvf}
F.~Arco, S.~Heinemeyer and M.J.~Herrero,  \textit{{Sensitivity to triple Higgs
  couplings via di-Higgs production in the 2HDM at $e^+e^-$ colliders}},
  \href{https://doi.org/10.1140/epjc/s10052-021-09665-w}{Eur. Phys. J. C
  {\bfseries 81} (2021) 913}
  [\href{https://arxiv.org/abs/2106.11105}{{\ttfamily 2106.11105}}].

\bibitem{Han:2021udl}
T.~Han, S.~Li, S.~Su, W.~Su and Y.~Wu,  \textit{{Heavy Higgs bosons in 2HDM at
  a muon collider}}, \href{https://doi.org/10.1103/PhysRevD.104.055029}{Phys.
  Rev. D {\bfseries 104} (2021) 055029}
  [\href{https://arxiv.org/abs/2102.08386}{{\ttfamily 2102.08386}}].

\bibitem{Li:2020hao}
S.~Li, H.~Song and S.~Su,  \textit{{Probing Exotic Charged Higgs Decays in the
  Type-II 2HDM through Top Rich Signal at a Future 100 TeV pp Collider}},
  \href{https://doi.org/10.1007/JHEP11(2020)105}{JHEP {\bfseries 11} (2020)
  105} [\href{https://arxiv.org/abs/2005.00576}{{\ttfamily 2005.00576}}].

\bibitem{Liu:2020kxt}
Y.-B.~Liu and S.~Moretti,  \textit{{Probing the top-Higgs boson FCNC couplings
  via the $h\to \gamma\gamma$ channel at the HE-LHC and FCC-hh}},
  \href{https://doi.org/10.1103/PhysRevD.101.075029}{Phys. Rev. D {\bfseries
  101} (2020) 075029} [\href{https://arxiv.org/abs/2002.05311}{{\ttfamily
  2002.05311}}].

\bibitem{Chung:2022kjp}
Y.-L.~Chung, K.~Cheung and S.-C.~Hsu,  \textit{{Sensitivity of
  two-Higgs-doublet models on Higgs-pair production via
  bb\textasciimacron{}bb\textasciimacron{} final state}},
  \href{https://doi.org/10.1103/PhysRevD.106.095015}{Phys. Rev. D {\bfseries
  106} (2022) 095015} [\href{https://arxiv.org/abs/2207.09602}{{\ttfamily
  2207.09602}}].

\bibitem{Bahl:2020kwe}
H.~Bahl, P.~Bechtle, S.~Heinemeyer, S.~Liebler, T.~Stefaniak and G.~Weiglein,
  \textit{{HL-LHC and ILC sensitivities in the hunt for heavy Higgs bosons}},
  \href{https://doi.org/10.1140/epjc/s10052-020-08472-z}{Eur. Phys. J. C
  {\bfseries 80} (2020) 916}
  [\href{https://arxiv.org/abs/2005.14536}{{\ttfamily 2005.14536}}].

\bibitem{Aoki:2009ha}
M.~Aoki, S.~Kanemura, K.~Tsumura and K.~Yagyu,  \textit{{Models of Yukawa
  interaction in the two Higgs doublet model, and their collider
  phenomenology}}, \href{https://doi.org/10.1103/PhysRevD.80.015017}{Phys. Rev.
  D {\bfseries 80} (2009) 015017}
  [\href{https://arxiv.org/abs/0902.4665}{{\ttfamily 0902.4665}}].

\bibitem{Ginzburg:2004vp}
I.F.~Ginzburg and M.~Krawczyk,  \textit{{Symmetries of two Higgs doublet model
  and CP violation}}, \href{https://doi.org/10.1103/PhysRevD.72.115013}{Phys.
  Rev. D {\bfseries 72} (2005) 115013}
  [\href{https://arxiv.org/abs/hep-ph/0408011}{{\ttfamily hep-ph/0408011}}].

\bibitem{Barroso:2013awa}
A.~Barroso, P.M.~Ferreira, I.P.~Ivanov and R.~Santos,  \textit{{Metastability
  bounds on the two Higgs doublet model}},
  \href{https://doi.org/10.1007/JHEP06(2013)045}{JHEP {\bfseries 06} (2013)
  045} [\href{https://arxiv.org/abs/1303.5098}{{\ttfamily 1303.5098}}].

\bibitem{Aoki:2021oez}
M.~Aoki, T.~Komatsu and H.~Shibuya,  \textit{{Possibility of a multi-step
  electroweak phase transition in the two-Higgs doublet models}},
  \href{https://doi.org/10.1093/ptep/ptac068}{PTEP {\bfseries 2022} (2022)
  063B05} [\href{https://arxiv.org/abs/2106.03439}{{\ttfamily 2106.03439}}].

\bibitem{Chakraborty:2015raa}
I.~Chakraborty and A.~Kundu,  \textit{{Scalar potential of two-Higgs doublet
  models}}, \href{https://doi.org/10.1103/PhysRevD.92.095023}{Phys. Rev. D
  {\bfseries 92} (2015) 095023}
  [\href{https://arxiv.org/abs/1508.00702}{{\ttfamily 1508.00702}}].

\bibitem{ParticleDataGroup:2024}
{\scshape Particle Data Group} collaboration,  \textit{{Review of particle
  physics}}, \href{https://doi.org/10.1103/PhysRevD.110.030001}{Phys. Rev. D
  {\bfseries 110} (2024) 030001}.

\bibitem{Ginzburg:2005dt}
I.F.~Ginzburg and I.P.~Ivanov,  \textit{{Tree-level unitarity constraints in
  the most general 2HDM}},
  \href{https://doi.org/10.1103/PhysRevD.72.115010}{Phys. Rev. D {\bfseries 72}
  (2005) 115010} [\href{https://arxiv.org/abs/hep-ph/0508020}{{\ttfamily
  hep-ph/0508020}}].

\bibitem{Peskin:1991sw}
M.E.~Peskin and T.~Takeuchi,  \textit{{Estimation of oblique electroweak
  corrections}}, \href{https://doi.org/10.1103/PhysRevD.46.381}{Phys. Rev. D
  {\bfseries 46} (1992) 381}.

\bibitem{Eriksson:2009ws}
D.~Eriksson, J.~Rathsman and O.~Stal,  \textit{{2HDMC: Two-Higgs-Doublet Model
  Calculator Physics and Manual}},
  \href{https://doi.org/10.1016/j.cpc.2009.09.011}{Comput. Phys. Commun.
  {\bfseries 181} (2010) 189}
  [\href{https://arxiv.org/abs/0902.0851}{{\ttfamily 0902.0851}}].

\bibitem{Mahmoudi:2007vz}
F.~Mahmoudi,  \textit{{SuperIso: A Program for calculating the isospin
  asymmetry of B ---\ensuremath{>} K* gamma in the MSSM}},
  \href{https://doi.org/10.1016/j.cpc.2007.12.006}{Comput. Phys. Commun.
  {\bfseries 178} (2008) 745}
  [\href{https://arxiv.org/abs/0710.2067}{{\ttfamily 0710.2067}}].

\bibitem{Mahmoudi:2008tp}
F.~Mahmoudi,  \textit{{SuperIso v2.3: A Program for calculating flavor physics
  observables in Supersymmetry}},
  \href{https://doi.org/10.1016/j.cpc.2009.02.017}{Comput. Phys. Commun.
  {\bfseries 180} (2009) 1579}
  [\href{https://arxiv.org/abs/0808.3144}{{\ttfamily 0808.3144}}].

\bibitem{Mahmoudi:2009zz}
F.~Mahmoudi,  \textit{{SuperIso v3.0, flavor physics observables calculations:
  Extension to NMSSM}},
  \href{https://doi.org/10.1016/j.cpc.2009.05.001}{Comput. Phys. Commun.
  {\bfseries 180} (2009) 1718}.

\bibitem{Bechtle:2008jh}
P.~Bechtle, O.~Brein, S.~Heinemeyer, G.~Weiglein and K.E.~Williams,
  \textit{{HiggsBounds: Confronting Arbitrary Higgs Sectors with Exclusion
  Bounds from LEP and the Tevatron}},
  \href{https://doi.org/10.1016/j.cpc.2009.09.003}{Comput. Phys. Commun.
  {\bfseries 181} (2010) 138}
  [\href{https://arxiv.org/abs/0811.4169}{{\ttfamily 0811.4169}}].

\bibitem{Bechtle:2011sb}
P.~Bechtle, O.~Brein, S.~Heinemeyer, G.~Weiglein and K.E.~Williams,
  \textit{{HiggsBounds 2.0.0: Confronting Neutral and Charged Higgs Sector
  Predictions with Exclusion Bounds from LEP and the Tevatron}},
  \href{https://doi.org/10.1016/j.cpc.2011.07.015}{Comput. Phys. Commun.
  {\bfseries 182} (2011) 2605}
  [\href{https://arxiv.org/abs/1102.1898}{{\ttfamily 1102.1898}}].

\bibitem{Bechtle:2012lvg}
P.~Bechtle, O.~Brein, S.~Heinemeyer, O.~Stal, T.~Stefaniak, G.~Weiglein et~al.,
   \textit{{Recent Developments in HiggsBounds and a Preview of HiggsSignals}},
  \href{https://doi.org/10.22323/1.156.0024}{PoS {\bfseries CHARGED2012} (2012)
  024} [\href{https://arxiv.org/abs/1301.2345}{{\ttfamily 1301.2345}}].

\bibitem{Bechtle:2013wla}
P.~Bechtle, O.~Brein, S.~Heinemeyer, O.~St\r{a}l, T.~Stefaniak, G.~Weiglein
  et~al.,  \textit{{$\mathsf{HiggsBounds}-4$: Improved Tests of Extended Higgs
  Sectors against Exclusion Bounds from LEP, the Tevatron and the LHC}},
  \href{https://doi.org/10.1140/epjc/s10052-013-2693-2}{Eur. Phys. J. C
  {\bfseries 74} (2014) 2693}
  [\href{https://arxiv.org/abs/1311.0055}{{\ttfamily 1311.0055}}].

\bibitem{Bechtle:2020pkv}
P.~Bechtle, D.~Dercks, S.~Heinemeyer, T.~Klingl, T.~Stefaniak, G.~Weiglein
  et~al.,  \textit{{HiggsBounds-5: Testing Higgs Sectors in the LHC 13 TeV
  Era}}, \href{https://doi.org/10.1140/epjc/s10052-020-08557-9}{Eur. Phys. J. C
  {\bfseries 80} (2020) 1211}
  [\href{https://arxiv.org/abs/2006.06007}{{\ttfamily 2006.06007}}].

\bibitem{Bahl:2021yhk}
H.~Bahl, V.M.~Lozano, T.~Stefaniak and J.~Wittbrodt,  \textit{{Testing exotic
  scalars with HiggsBounds}},
  \href{https://doi.org/10.1140/epjc/s10052-022-10446-2}{Eur. Phys. J. C
  {\bfseries 82} (2022) 584}
  [\href{https://arxiv.org/abs/2109.10366}{{\ttfamily 2109.10366}}].

\bibitem{Bahl:2022igd}
H.~Bahl, T.~Biek\"otter, S.~Heinemeyer, C.~Li, S.~Paasch, G.~Weiglein et~al.,
  \textit{{HiggsTools: BSM scalar phenomenology with new versions of
  HiggsBounds and HiggsSignals}},
  \href{https://doi.org/10.1016/j.cpc.2023.108803}{Comput. Phys. Commun.
  {\bfseries 291} (2023) 108803}
  [\href{https://arxiv.org/abs/2210.09332}{{\ttfamily 2210.09332}}].

\bibitem{HeavyFlavorAveragingGroup:2022wzx}
{\scshape Heavy Flavor Averaging Group, HFLAV} collaboration,
  \textit{{Averages of b-hadron, c-hadron, and \ensuremath{\tau}-lepton
  properties as of 2021}},
  \href{https://doi.org/10.1103/PhysRevD.107.052008}{Phys. Rev. D {\bfseries
  107} (2023) 052008} [\href{https://arxiv.org/abs/2206.07501}{{\ttfamily
  2206.07501}}].

\bibitem{LHCb:2021vsc}
{\scshape LHCb} collaboration,  \textit{{Analysis of Neutral B-Meson Decays
  into Two Muons}}, \href{https://doi.org/10.1103/PhysRevLett.128.041801}{Phys.
  Rev. Lett. {\bfseries 128} (2022) 041801}
  [\href{https://arxiv.org/abs/2108.09284}{{\ttfamily 2108.09284}}].

\bibitem{Huber:2015sra}
T.~Huber, T.~Hurth and E.~Lunghi,  \textit{{Inclusive $ \overline{B}\to
  {X}_s{\ell}^{+}{\ell}^{-} $ : complete angular analysis and a thorough study
  of collinear photons}}, \href{https://doi.org/10.1007/JHEP06(2015)176}{JHEP
  {\bfseries 06} (2015) 176}
  [\href{https://arxiv.org/abs/1503.04849}{{\ttfamily 1503.04849}}].

\bibitem{Belle-II:2018jsg}
{\scshape Belle-II} collaboration,  \textit{{The Belle II Physics Book}},
  \href{https://doi.org/10.1093/ptep/ptz106}{PTEP {\bfseries 2019} (2019)
  123C01} [\href{https://arxiv.org/abs/1808.10567}{{\ttfamily 1808.10567}}].

\bibitem{Huber:2024rbw}
T.~Huber, T.~Hurth, J.~Jenkins, E.~Lunghi, Q.~Qin and K.K.~Vos,
  \textit{{Inclusive $ \overline{B}\to {X}_s{\ell}^{+}{\ell}^{-} $ at the LHC:
  theory predictions and new-physics reach}},
  \href{https://doi.org/10.1007/JHEP11(2024)130}{JHEP {\bfseries 11} (2024)
  130} [\href{https://arxiv.org/abs/2404.03517}{{\ttfamily 2404.03517}}].

\bibitem{LHCb:2016ykl}
{\scshape LHCb} collaboration,  \textit{{Measurements of the S-wave fraction in
  $B^{0}\rightarrow K^{+}\pi^{-}\mu^{+}\mu^{-}$ decays and the
  $B^{0}\rightarrow K^{\ast}(892)^{0}\mu^{+}\mu^{-}$ differential branching
  fraction}}, \href{https://doi.org/10.1007/JHEP11(2016)047}{JHEP {\bfseries
  11} (2016) 047} [\href{https://arxiv.org/abs/1606.04731}{{\ttfamily
  1606.04731}}].

\bibitem{Bechtle:2013xfa}
P.~Bechtle, S.~Heinemeyer, O.~St\r{a}l, T.~Stefaniak and G.~Weiglein,
  \textit{{$HiggsSignals$: Confronting arbitrary Higgs sectors with
  measurements at the Tevatron and the LHC}},
  \href{https://doi.org/10.1140/epjc/s10052-013-2711-4}{Eur. Phys. J. C
  {\bfseries 74} (2014) 2711}
  [\href{https://arxiv.org/abs/1305.1933}{{\ttfamily 1305.1933}}].

\bibitem{Stal:2013hwa}
O.~St\r{a}l and T.~Stefaniak,  \textit{{Constraining extended Higgs sectors
  with HiggsSignals}}, \href{https://doi.org/10.22323/1.180.0314}{PoS
  {\bfseries EPS-HEP2013} (2013) 314}
  [\href{https://arxiv.org/abs/1310.4039}{{\ttfamily 1310.4039}}].

\bibitem{Bechtle:2014ewa}
P.~Bechtle, S.~Heinemeyer, O.~St\r{a}l, T.~Stefaniak and G.~Weiglein,
  \textit{{Probing the Standard Model with Higgs signal rates from the
  Tevatron, the LHC and a future ILC}},
  \href{https://doi.org/10.1007/JHEP11(2014)039}{JHEP {\bfseries 11} (2014)
  039} [\href{https://arxiv.org/abs/1403.1582}{{\ttfamily 1403.1582}}].

\bibitem{Bechtle:2020uwn}
P.~Bechtle, S.~Heinemeyer, T.~Klingl, T.~Stefaniak, G.~Weiglein and
  J.~Wittbrodt,  \textit{{HiggsSignals-2: Probing new physics with precision
  Higgs measurements in the LHC 13 TeV era}},
  \href{https://doi.org/10.1140/epjc/s10052-021-08942-y}{Eur. Phys. J. C
  {\bfseries 81} (2021) 145}
  [\href{https://arxiv.org/abs/2012.09197}{{\ttfamily 2012.09197}}].

\bibitem{Feroz:2007kg}
F.~Feroz and M.P.~Hobson,  \textit{{Multimodal nested sampling: an efficient
  and robust alternative to MCMC methods for astronomical data analysis}},
  \href{https://doi.org/10.1111/j.1365-2966.2007.12353.x}{Mon. Not. Roy.
  Astron. Soc. {\bfseries 384} (2008) 449}
  [\href{https://arxiv.org/abs/0704.3704}{{\ttfamily 0704.3704}}].

\bibitem{Feroz:2008xx}
F.~Feroz, M.P.~Hobson and M.~Bridges,  \textit{{MultiNest: an efficient and
  robust Bayesian inference tool for cosmology and particle physics}},
  \href{https://doi.org/10.1111/j.1365-2966.2009.14548.x}{Mon. Not. Roy.
  Astron. Soc. {\bfseries 398} (2009) 1601}
  [\href{https://arxiv.org/abs/0809.3437}{{\ttfamily 0809.3437}}].

\bibitem{Feroz:2013hea}
F.~Feroz, M.P.~Hobson, E.~Cameron and A.N.~Pettitt,  \textit{{Importance Nested
  Sampling and the MultiNest Algorithm}},
  \href{https://doi.org/10.21105/astro.1306.2144}{Open J. Astrophys. {\bfseries
  2} (2019) 10} [\href{https://arxiv.org/abs/1306.2144}{{\ttfamily
  1306.2144}}].

\bibitem{Nelder:1965zz}
J.A.~Nelder and R.~Mead,  \textit{{A Simplex Method for Function
  Minimization}}, \href{https://doi.org/10.1093/comjnl/7.4.308}{Comput. J.
  {\bfseries 7} (1965) 308}.

\bibitem{10.2307/2346772}
R.~O'Neill,  \textit{Algorithm as 47: Function minimization using a simplex
  procedure}, {Journal of the Royal Statistical Society. Series C (Applied
  Statistics) {\bfseries 20} (1971) 338}.

\bibitem{Enomoto:2015wbn}
T.~Enomoto and R.~Watanabe,  \textit{{Flavor constraints on the Two Higgs
  Doublet Models of Z$_{2}$ symmetric and aligned types}},
  \href{https://doi.org/10.1007/JHEP05(2016)002}{JHEP {\bfseries 05} (2016)
  002} [\href{https://arxiv.org/abs/1511.05066}{{\ttfamily 1511.05066}}].

\bibitem{Grimus:2008nb}
W.~Grimus, L.~Lavoura, O.M.~Ogreid and P.~Osland,  \textit{{The Oblique
  parameters in multi-Higgs-doublet models}},
  \href{https://doi.org/10.1016/j.nuclphysb.2008.04.019}{Nucl. Phys. B
  {\bfseries 801} (2008) 81} [\href{https://arxiv.org/abs/0802.4353}{{\ttfamily
  0802.4353}}].

\bibitem{Misiak:2020vlo}
M.~Misiak, A.~Rehman and M.~Steinhauser,  \textit{{Towards $ \overline{B}\to
  {X}_s\gamma $ at the NNLO in QCD without interpolation in m$_{c}$}},
  \href{https://doi.org/10.1007/JHEP06(2020)175}{JHEP {\bfseries 06} (2020)
  175} [\href{https://arxiv.org/abs/2002.01548}{{\ttfamily 2002.01548}}].

\bibitem{Charles:2004jd}
{\scshape CKMfitter Group} collaboration,  \textit{{CP violation and the CKM
  matrix: Assessing the impact of the asymmetric $B$ factories}},
  \href{https://doi.org/10.1140/epjc/s2005-02169-1}{Eur. Phys. J. C {\bfseries
  41} (2005) 1} [\href{https://arxiv.org/abs/hep-ph/0406184}{{\ttfamily
  hep-ph/0406184}}].

\bibitem{CKMFitterLink}
{\scshape CKMfitter Group} collaboration,
  \textit{{http://ckmfitter.in2p3.fr/www/results/plots\_summer23/num/ckmEval\_results\_summer23.html}},
  .

\bibitem{Chen:2006nua}
C.-H.~Chen and C.-Q.~Geng,  \textit{{Charged Higgs on B- ---\ensuremath{>} tau
  anti-nu(tau) and anti-B ---\ensuremath{>} P(V) l anti-nu(l)}},
  \href{https://doi.org/10.1088/1126-6708/2006/10/053}{JHEP {\bfseries 10}
  (2006) 053} [\href{https://arxiv.org/abs/hep-ph/0608166}{{\ttfamily
  hep-ph/0608166}}].

\bibitem{Descotes-Genon:2015uva}
S.~Descotes-Genon, L.~Hofer, J.~Matias and J.~Virto,  \textit{{Global analysis
  of $b\to s\ell\ell$ anomalies}},
  \href{https://doi.org/10.1007/JHEP06(2016)092}{JHEP {\bfseries 06} (2016)
  092} [\href{https://arxiv.org/abs/1510.04239}{{\ttfamily 1510.04239}}].

\bibitem{LHCHiggsCrossSectionWorkingGroup:2016ypw}
{\scshape LHC Higgs Cross Section Working Group} collaboration,
  \textit{{Handbook of LHC Higgs Cross Sections: 4. Deciphering the Nature of
  the Higgs Sector}},  \href{https://arxiv.org/abs/1610.07922}{{\ttfamily
  1610.07922}}.

\bibitem{LHCHiggsCrossSectionWorkingGroup:2013rie}
{\scshape LHC Higgs Cross Section Working Group} collaboration,
  \textit{{Handbook of LHC Higgs Cross Sections: 3. Higgs Properties}},
  \href{https://arxiv.org/abs/1307.1347}{{\ttfamily 1307.1347}}.

\bibitem{Krause:2018wmo}
M.~Krause, M.~M\"uhlleitner and M.~Spira,  \textit{{2HDECAY \textemdash{}A
  program for the calculation of electroweak one-loop corrections to Higgs
  decays in the Two-Higgs-Doublet Model including state-of-the-art QCD
  corrections}}, \href{https://doi.org/10.1016/j.cpc.2019.08.003}{Comput. Phys.
  Commun. {\bfseries 246} (2020) 106852}
  [\href{https://arxiv.org/abs/1810.00768}{{\ttfamily 1810.00768}}].

\bibitem{Djouadi:1997yw}
A.~Djouadi, J.~Kalinowski and M.~Spira,  \textit{{HDECAY: A Program for Higgs
  boson decays in the standard model and its supersymmetric extension}},
  \href{https://doi.org/10.1016/S0010-4655(97)00123-9}{Comput. Phys. Commun.
  {\bfseries 108} (1998) 56}
  [\href{https://arxiv.org/abs/hep-ph/9704448}{{\ttfamily hep-ph/9704448}}].

\bibitem{Djouadi:2018xqq}
{\scshape HDECAY} collaboration,  \textit{{HDECAY: Twenty$_{++}$ years after}},
  \href{https://doi.org/10.1016/j.cpc.2018.12.010}{Comput. Phys. Commun.
  {\bfseries 238} (2019) 214}
  [\href{https://arxiv.org/abs/1801.09506}{{\ttfamily 1801.09506}}].

\bibitem{Krause:2016oke}
M.~Krause, R.~Lorenz, M.~Muhlleitner, R.~Santos and H.~Ziesche,
  \textit{{Gauge-independent Renormalization of the 2-Higgs-Doublet Model}},
  \href{https://doi.org/10.1007/JHEP09(2016)143}{JHEP {\bfseries 09} (2016)
  143} [\href{https://arxiv.org/abs/1605.04853}{{\ttfamily 1605.04853}}].

\bibitem{Krause:2016xku}
M.~Krause, M.~Muhlleitner, R.~Santos and H.~Ziesche,  \textit{{Higgs-to-Higgs
  boson decays in a 2HDM at next-to-leading order}},
  \href{https://doi.org/10.1103/PhysRevD.95.075019}{Phys. Rev. D {\bfseries 95}
  (2017) 075019} [\href{https://arxiv.org/abs/1609.04185}{{\ttfamily
  1609.04185}}].

\bibitem{Denner:2018opp}
A.~Denner, S.~Dittmaier and J.-N.~Lang,  \textit{{Renormalization of mixing
  angles}}, \href{https://doi.org/10.1007/JHEP11(2018)104}{JHEP {\bfseries 11}
  (2018) 104} [\href{https://arxiv.org/abs/1808.03466}{{\ttfamily
  1808.03466}}].

\bibitem{Hahn:1998yk}
T.~Hahn and M.~Perez-Victoria,  \textit{{Automatized one loop calculations in
  four-dimensions and D-dimensions}},
  \href{https://doi.org/10.1016/S0010-4655(98)00173-8}{Comput. Phys. Commun.
  {\bfseries 118} (1999) 153}
  [\href{https://arxiv.org/abs/hep-ph/9807565}{{\ttfamily hep-ph/9807565}}].

\bibitem{Staub:2008uz}
F.~Staub,  \textit{{SARAH}},  \href{https://arxiv.org/abs/0806.0538}{{\ttfamily
  0806.0538}}.

\bibitem{Staub:2013tta}
F.~Staub,  \textit{{SARAH 4 : A tool for (not only SUSY) model builders}},
  \href{https://doi.org/10.1016/j.cpc.2014.02.018}{Comput. Phys. Commun.
  {\bfseries 185} (2014) 1773}
  [\href{https://arxiv.org/abs/1309.7223}{{\ttfamily 1309.7223}}].

\bibitem{Staub:2009bi}
F.~Staub,  \textit{{From Superpotential to Model Files for FeynArts and
  CalcHep/CompHep}}, \href{https://doi.org/10.1016/j.cpc.2010.01.011}{Comput.
  Phys. Commun. {\bfseries 181} (2010) 1077}
  [\href{https://arxiv.org/abs/0909.2863}{{\ttfamily 0909.2863}}].

\bibitem{Boos:1994xb}
E.E.~Boos, M.N.~Dubinin, V.A.~Ilyin, A.E.~Pukhov and V.I.~Savrin,
  \textit{{CompHEP: Specialized package for automatic calculations of
  elementary particle decays and collisions}},
  \href{https://arxiv.org/abs/hep-ph/9503280}{{\ttfamily hep-ph/9503280}}.

\bibitem{Pukhov:1999gg}
A.~Pukhov, E.~Boos, M.~Dubinin, V.~Edneral, V.~Ilyin, D.~Kovalenko et~al.,
  \textit{{CompHEP: A Package for evaluation of Feynman diagrams and
  integration over multiparticle phase space}},
  \href{https://arxiv.org/abs/hep-ph/9908288}{{\ttfamily hep-ph/9908288}}.

\bibitem{Pukhov:2004ca}
A.~Pukhov,  \textit{{CalcHEP 2.3: MSSM, structure functions, event generation,
  batchs, and generation of matrix elements for other packages}},
  \href{https://arxiv.org/abs/hep-ph/0412191}{{\ttfamily hep-ph/0412191}}.

\bibitem{Yuasa:1999rg}
F.~Yuasa et~al.,  \textit{{Automatic computation of cross-sections in HEP:
  Status of GRACE system}}, \href{https://doi.org/10.1143/PTPS.138.18}{Prog.
  Theor. Phys. Suppl. {\bfseries 138} (2000) 18}
  [\href{https://arxiv.org/abs/hep-ph/0007053}{{\ttfamily hep-ph/0007053}}].

\bibitem{Fujimoto:2002sj}
J.~Fujimoto et~al.,  \textit{{GRACE/SUSY automatic generation of tree
  amplitudes in the minimal supersymmetric standard model}},
  \href{https://doi.org/10.1016/S0010-4655(03)00159-0}{Comput. Phys. Commun.
  {\bfseries 153} (2003) 106}
  [\href{https://arxiv.org/abs/hep-ph/0208036}{{\ttfamily hep-ph/0208036}}].

\bibitem{Grace:2006}
{\scshape Minami-Tateya} collaboration, ``{GRACE} version 2.2.1.''
  \url{http://minami-home.kek.jp/}.

\bibitem{Behnke:2013xla}
T.~Behnke, J.E.~Brau, B.~Foster, J.~Fuster, M.~Harrison, J.M.~Paterson et~al.,
  \textit{{The International Linear Collider Technical Design Report - Volume
  1: Executive Summary}},  \href{https://arxiv.org/abs/1306.6327}{{\ttfamily
  1306.6327}}.

\bibitem{Fujii:2013lba}
K.~Fujii,  \textit{{Physics at the ILC with focus mostly on Higgs physics}},
  in \emph{{1st Toyama International Workshop on Higgs as a Probe of New
  Physics 2013}}, 5, 2013 [\href{https://arxiv.org/abs/1305.1692}{{\ttfamily
  1305.1692}}].

\bibitem{Fujimoto:1990tb}
J.~Fujimoto, M.~Igarashi, N.~Nakazawa, Y.~Shimizu and K.~Tobimatsu,
  \textit{{Radiative corrections to $e^{+} e^{-}$ reactions in electroweak
  theory}}, \href{https://doi.org/10.1143/PTPS.100.1}{Prog. Theor. Phys. Suppl.
  {\bfseries 100} (1990) 1}.

\end{thebibliography}\endgroup
\end{document}